\documentclass[3p,10pt]{my_elsarticle}
\usepackage{etoolbox}
\newtoggle{submission}
\toggletrue{submission}
\biboptions{longnamesfirst,square,sort,compress}
\usepackage[nodots]{numcompress}
\journal{Composite Structures}
\usepackage[utf8]{inputenc}
\usepackage[T1]{fontenc}
\usepackage{graphicx}
\usepackage{enumerate}
\usepackage{color,amssymb,amsmath}
\usepackage{empheq}
\usepackage{mathtools}
\usepackage{mathrsfs} 
\usepackage{frcursive}
\usepackage{comment}
\usepackage{multirow}
\usepackage{booktabs}
\usepackage{tikz}
\usepackage[font=footnotesize,labelfont=bf,tableposition=top]{caption}

\usepackage{lscape}
\usepackage[ruled,vlined]{algorithm2e}

\iftoggle{submission}{}{
  
  \usepackage{pgfplots}
  \usetikzlibrary{pgfplots.groupplots}
  \usetikzlibrary{calc}
  \pgfplotsset{compat=1.7}
  \tikzset{every mark/.append style={scale=0.5}}
  \usepgfplotslibrary{external}
  \tikzset{external/force remake}
  \tikzexternalize
}

\newcommand{\fozz}{FoZZ\nobreakdash-4}
\newcommand{\tozz}{ToZZ\nobreakdash-4}
\newcommand{\tdwf}{3D\nobreakdash-WF}
\newcommand{\itwf}{Ite\nobreakdash-WF}

\newlength{\GraphWidth}
\newlength{\GraphHeight}
\newlength{\GraphHorizSep}
\newlength{\GraphVerticSep}
\newlength{\GraphVerticLegendSep}
\newlength{\HorizVerticLegendSep}
\makeatletter
\if@twocolumn
  \newenvironment{SmartFigure}{\begin{figure}}{\end{figure}}
\else
  \newenvironment{SmartFigure}{\begin{figure*}}{\end{figure*}}
\fi
\makeatother

\usepackage{float}
\restylefloat{table}

\newcommand{\mymat}[1]{
[\mathbf{#1}]}
\newcommand{\myvect}[1]{
\{\mathbf{#1}\}}

\input cyracc.def

\everymath{\displaystyle}

\usepackage{hyperref}
\hypersetup{
    unicode=true,          
    pdftoolbar=true,        
    pdfmenubar=true,        
    pdffitwindow=true,     
    pdfstartview={FitH},    
    pdftitle={Two multilayered plate models with transverse shear warping functions issued from three dimensional elasticity equations},    
    pdfauthor={Alexandre Loredo, Alexis Castel},     
    pdfcreator={Alexandre Loredo},   
    pdfproducer={Alexandre Loredo}, 
    pdfnewwindow=true,      
    colorlinks=true,       
    linkcolor=red,          
    citecolor=green,        
    filecolor=magenta,      
    urlcolor=cyan           
}
\usepackage[shortcuts]{extdash}
\begin{document}
\begin{frontmatter}
  \title{Two multilayered plate models with transverse shear warping functions issued from three dimensional elasticity equations}
    \author[drive]{A.~Loredo\corref{cor1}}
  	\ead{alexandre.loredo@u-bourgogne.fr}
  	\author[drive]{A.~Castel}
  	\ead{alexis.castel@u-bourgogne.fr}
  	\cortext[cor1]{Corresponding author}
  	\address[drive]{DRIVE, Universit\'e de Bourgogne, 49 rue Mlle Bourgeois, 58027 Nevers, France}
	\begin{abstract}
A multilayered plate theory which uses transverse shear warping functions is presented. Two methods to obtain the transverse shear warping functions from three-dimensional elasticity equations are proposed. The warping functions are issued from the variations of transverse shear stresses computed at specific points of a simply supported plate. The first method considers an exact 3D solution of the problem. The second method uses the solution provided by the model itself: the transverse shear stresses are computed integrating equilibrium equations. Hence, an iterative process is applied, the model is updated with the new warping functions, and so on. Once the sets of warping functions are obtained, the stiffness and mass matrices of the models are computed. These two models are compared to other models and to analytical solutions for the bending of simply supported plates. Four different laminates and a sandwich plate are considered. Their length-to-thickness ratios vary from 2 to 100. An additional analytical solution that simulates the behavior of laminates under the plane stress hypothesis -- shared by all the considered models -- is computed. Both presented models give results very close to this exact solution, for all laminates and all length-to-thickness ratios.
	\end{abstract}
\end{frontmatter}
\section{Introduction}
In many human-built structures, plates and shells are present. These particular structures are distinguished from others because a dimension -- the transverse dimension -- is much smaller than the others. Hence, although it is always possible, their representation through a three-dimensional domain is not the best way to study them. To understand and predict their mechanical behavior, plate models have been developed. These models permit to study plates and shells through a two-dimensional domain while allowing at least membrane and bending deformations. What happens in the third direction is not ignored, it is precisely the purpose of the plate model to integrate the transverse behavior into its equations. The better this behavior is described, the more accurate the model will be. History of plate models probably begins with works of Kirchhoff, Love, and  Rayleigh~\cite{Kirchhoff1850,Love1888,Rayleigh1944} leading to the Love-Kirchhoff model in which no shear deformation is allowed. Because of the limitation of this model to thin plates, authors like Reissner, Hencky, Bolle, Uflyand, and Mindlin\cite{Reissner1945, Hencky1947, Bolle1947, Uflyand1948, Mindlin1951} have proposed to integrate the shear phenomenon into their formulation. In particular, Reissner made the assumption that shear stresses have a parabolic distribution and consider the normal stress in his model which is derived from a complementary energy. On the other hand, Hencky and Mindlin considered a linear variation of the displacements $u$ and $v$ and no transverse strain. Mindlin and Reisnner theories are often associated but it is incorrect as demonstrated in reference~\cite{Wang2001}. The Love-Kirchhoff and the Hencky--Mindlin models were first proposed for homogeneous plates but their pendent for laminated structures have been later presented and are called the Classical Laminated Plate Theory (CLPT) and the First-order Shear Deformation laminated plate Theory (FoSDT). The last one has been improved by the use of shear correction factors~\cite{Whitney1973,Noor1989,Pai1995}. The need to improve this accuracy has been motivated by the study of thick plates and laminated plates. In both cases, the early plate models fail to give accurate results. It is even worse if some layers have low mechanical properties compared to others, which is the case for sandwich structures or when viscoelastic layers are used to improve the damping. To overcome these difficulties, specific models have been proposed. However, a universal model that can manage plates of various length-to-thickness ratios, with any lamination scheme and various materials including functionally graded materials, with good accuracy for the static, dynamic, and damped dynamic behaviors is still a challenge.
\par
Dealing with multilayered plates has given rise to another class of theories, called the Layer-Wise (LW) models, in which the number of unknowns depends on the number of layers, by opposition to the Equivalent Single Layer (ESL) family of models in which the number of unknowns is independent of the number of layers. Obviously, LW models are expected to be more accurate than ESL models but they are less easy to implement for complex structures and require more computational resources than ESL models. This has motivated researchers to propose ESL models in which the transverse shear is taken into account in a more precise way than for the first models. With the help of the classification given by Carrera~\cite{Carrera2002} among other review papers, we can distinguish different approaches that have given interesting ESL models:
\begin{itemize}
  \item Higher-order (and also non-polynomial) shear deformation theories: the Vlasov--Levinson--Reddy~\cite{Vlasov1957,Levinson1980,Reddy1984} theory, also called Third order Shear Deformation Theory (ToSDT), among other third order theories, proposes a kinematic field with a third-order polynomial dependence on $z$, motivated by the respect of the nullity of transverse shear at top and bottom faces of the plate. Higher order (order greater than 3) theories have been proposed. Non-polynomial functions have also been used to take into account the shear phenomenon: for example, Touratier in reference~\cite{Touratier1991}, uses a $f(z)=h/\pi\sin(\pi z/h)$ function, Soldatos~\cite{Soldatos1992} uses a $f(z)=z\cosh(1/2)-h\sinh(z/h)$ function, and Thai \& al.~\cite{Thai2014} uses a $f(z)=h\arctan(2z/h)-z$ function to integrate the transverse shear in the kinematic field. Note that, for the study of laminated plates, these models do not integrate information about the lamination sequence in their kinematic field, and do not verify the transverse stress continuity at interfaces.
  \item Interlaminar continuous ESL models: some \emph{a priori} LW models reduce to ESL models with the help of assumptions between the fields in each layer. Zig-Zag (ZZ) models enter in this category. Pioneering works of Lekhnitskii~\cite{Lekhnitskii1935} and Ambartsumyan~\cite{Ambartsumyan1958} have been classified as such by Carrera~\cite{Carrera2002} who also shows that other authors have integrated the multilayer structure in their model~\cite{Whitney1969, Sun1973, Cho1993} in a very similar manner. The main idea of ZZ models is to let in-plane displacements vary with $z$ according to the superposition of a zig-zag law to a global law -- cubic for example. With these models, shear stresses can satisfy both continuity at interfaces and null (or prescribed) values at top and bottom faces of the plate. An enhancement of the second model presented in reference~\cite{Sun1973} for general lamination sequences is proposed in references~\cite{Woodcock2008,Loredo2013}. Models of references~\cite{Loredo2013, Cho1993} are used for comparison in this study, they are presented at section~\ref{sec:OtherModels}. Other works can be cited. For example, Pai~\cite{Pai1995} presents a method to determine four warping functions of cubic order in each layer, and also a generic method to compute shear correction factors. In ref.~\cite{Arya2002}, the author proposes a warping function for beam problems which is a linear ZZ function superimposed to an overall sine function. In ref.~\cite{Kim2006}, the authors develop an enhancement of the model given in ref.~\cite{Cho1993} for general lamination sequences. Note that references~\cite{Loredo2013,Cho1993,Pai1995,Kim2006} propose models with four WF, which is necessary for the study of general lamination sequences. 
  \item Direct approaches and micropolar plate theories: based on the works of the Cosserat brothers~\cite{Cosserat1909} on generalized continuum mechanics and on deformable lines and surfaces. The micropolar theory of elasticity, which considers (microscopic) continuous rotations and couples in addition to the classical displacements and forces, has been applied to rods and shells by Ericksen~\cite{Ericksen1957}. Later, it has been used in numerous works, including plate models, as it can be seen in the review paper~\cite{Altenbach2010}.
  \item Sandwich theories: sandwich panels have been early considered apart from laminated composite because of their particular structure. They are made of two face layers (themselves made of one or several layers) and a core layer. The thickness of the faces is small compared to the thickness of the core (typically 10 times less) and Young modulus of the faces is high compared to Young modulus of the core (typically more than 1000 times higher). For these reasons, specific models have been proposed, for example by Reissner~\cite{Reissner1947}, followed by many others, as it can be seen in the review papers~\cite{Noor1996,Kreja2011}.
\end{itemize}
Additional references can be found in recent reviews on the subject~\cite{Wanji2008,Kreja2011,Khandan2012}.
\par
When theories are issued from assumptions on displacement or stress fields, there is no guarantee for these theories to be consistent. This means that, with respect to the required order of the considered theory, some terms of the three\-/dimensional strain energy may appear with an erroneous coefficient, or may even be omitted. The consistency of plate theories has been discussed by many authors, one can for example refer to recent works~\cite{Schneider2011,Schneider2014}. Precise rules for the consistency are formulated for homogeneous materials, but it seems difficult to extrapolate them for inhomogeneous materials like laminates and sandwiches, especially when advanced kinematic and/or static assumptions are made. Although this subject is not treated in this study, it can be a significant aspect as it is mentioned in the conclusion. 
\par
In this work, the kinematic assumptions are similar of those taken in Ref.~\cite{Kim2006} but they differ because the choice of the functions describing the transverse behavior is not made \emph{a priori}. These functions, called \emph{warping functions} (WF) are the core of the model. The model has been entirely formulated in~\cite{Loredo2013} and it has been shown that, according to specific choices of the WF, it can also represent most of the classical models (CLPT, FoSDT, ToSDT) and others, as it will appear in the following. However, and it is precisely the subject of this article, it is possible to choose and adapt the WF in a completely free manner.
\par
This article presents two different ways of obtaining new sets of WF issued from three dimensional elasticity laws. The first way consists in building the WF from three-dimensional solutions. Three-dimensional solutions for the bending of laminates have been first obtained by Pagano and by Srinivas \& al.~\cite{Pagano1970,Srinivas1970} for cross-ply laminates, by Noor \& Burton~\cite{Noor1990} for antisymmetric angle-ply laminates, and have been recently obtained for general lamination schemes~\cite{Loredo2014}. These solutions are achieved for particular boundary conditions and load, which can be seen as the principal limitation of their use in the present model. The second way is to derive the WF from the equilibrium equations. This leads to an iterative model: starting with ``classical'' WF, for example Reddy's formula, WF are issued from the equilibrium equations and then integrated to the model, and so on until no significant changes are detected.
\section{Considered plate theory}\label{sec:Formulation}
In this section, we recall the main components of the theory presented in Ref.~\cite{Loredo2013}. It is a plate theory based on the use of transverse shear WF. It allows the simulation of multilayer laminates made of orthotropic plies using different sets of transverse shear WF. The purpose of this paper is to propose enhanced WF issued from 3D elasticity equations (see section~\ref{sec:WF}), but several plate models (among ESL and ZZ models) issued from the literature can also be formulated in terms of transverse shear WF, as it was shown in Ref.~\cite{Loredo2013}. This is of practical interest when comparing results issued from different models, because these models can be implemented in a similar manner.
\subsection{Laminate definition and index convention}
The laminate, of height $h$, is composed of $N$ layers. All the quantities will be related to those of the middle plane\footnote{The reference plane can be arbitrarily chosen in the laminate assuming that the corresponding transverse shear WF are adapted in consequence.} which is placed at $z=0$; they are marked with the superscript $0$. 
In the following, Greek subscripts take values $1$ or $2$ and Latin subscripts take values $1$, $2$ or $3$. The Einstein's summation convention is used for subscripts only. The comma used as a subscript index means the partial derivative with respect to the following index(ices).
\subsection{Displacement field}\label{sec:DisplacementField}
The kinematic assumptions of the present theory are:
\begin{equation}
\label{eq:depl_L}
  \left\{
    \begin{array}{lll}
      u_\alpha(x,y,z) & = & u_\alpha^0(x,y) - z w^0_{,\alpha}(x,y) + \varphi_{\alpha\beta}(z)\gamma^0_{\beta3}(x,y) \\ 
      u_3(x,y,z) & = & w^0(x,y) 
    \end{array} 
  \right.
\end{equation}
where $u_\alpha^0(x,y)$, $w^0(x,y)$ and $\gamma^0_{\alpha3}(x,y)$, are respectively the in-plane displacements, the deflection and the engineering transverse shear strains evaluated at the middle plane. The $\varphi_{\alpha\beta}(z)$ are the four WF.
The associated strain field is derived from equation~\eqref{eq:depl_L}:
\makeatletter%
\if@twocolumn%
  \begin{subequations}\label{eq:strains}
  \begin{empheq}[left=\empheqlbrace]{align} \nonumber
    \varepsilon_{\alpha\beta}(x,y,z) & = \varepsilon_{\alpha\beta}^0(x,y) - z w^0_{,\alpha\beta}(x,y) + \frac{1}{2}\Big[\varphi_{\alpha\gamma}(z)\gamma^0_{\gamma3,\beta}(x,y) \\ 
           & \quad +\varphi_{\beta\gamma}(z)\gamma^0_{\gamma3,\alpha}(x,y)\Big] \label{eq:in_plane_strains}\\ 
    \varepsilon_{\alpha3}(x,y,z) & = \frac{1}{2}\varphi'_{\alpha\beta}(z)\gamma^0_{\beta3}(x,y) \label{eq:transverse_strains}\\ 
    \varepsilon_{33}(x,y,z) & = 0 \phantom{\frac{1}{2}}
  \end{empheq}
  \end{subequations}
\else%
  \begin{subequations}\label{eq:strains}
  \begin{empheq}[left=\empheqlbrace]{align}
    \varepsilon_{\alpha\beta}(x,y,z) & = \varepsilon_{\alpha\beta}^0(x,y) - z w^0_{,\alpha\beta}(x,y) 
           + \frac{1}{2}\left(\varphi_{\alpha\gamma}(z)\gamma^0_{\gamma3,\beta}(x,y)+\varphi_{\beta\gamma}(z)\gamma^0_{\gamma3,\alpha}(x,y)\right) \label{eq:in_plane_strains}\\ 
    \varepsilon_{\alpha3}(x,y,z) & = \frac{1}{2}\varphi'_{\alpha\beta}(z)\gamma^0_{\beta3}(x,y) \label{eq:transverse_strains}\\ 
    \varepsilon_{33}(x,y,z) & = 0 \phantom{\frac{1}{2}}
  \end{empheq}
  \end{subequations}
\fi%
\makeatother%
which, with the use of Hooke's law, leads to the following stress field:
\makeatletter%
\if@twocolumn%
  \begin{subequations}\label{eq:stresses}
  \begin{empheq}[left=\empheqlbrace]{align} \nonumber
    \sigma_{\alpha\beta}(x,y,z) &= Q_{\alpha\beta\gamma\delta}(z)\Big[\varepsilon_{\gamma\delta}^0(x,y) - z w^0_{,\gamma\delta}(x,y) \\
           & \quad + \varphi_{\gamma\mu}(z)\gamma^0_{\mu3,\delta}(x,y)\Big] \label{eq:in_plane_stresses}\\ 
    \sigma_{\alpha3}(x,y,z) &= C_{\alpha3\beta3}(z)\varphi'_{\beta\mu}(z)\gamma^0_{\mu3}(x,y) \label{eq:transverse_stresses2}\\ 
    \sigma_{33}(x,y,z) &= 0
  \end{empheq}
  \end{subequations}
\else%
  \begin{subequations}\label{eq:stresses}
  \begin{empheq}[left=\empheqlbrace]{align}
    \sigma_{\alpha\beta}(x,y,z) & = Q_{\alpha\beta\gamma\delta}(z)\left(\varepsilon_{\gamma\delta}^0(x,y) - z w^0_{,\gamma\delta}(x,y)
           + \varphi_{\gamma\mu}(z)\gamma^0_{\mu3,\delta}(x,y)\right) \label{eq:in_plane_stresses}\\ 
    \sigma_{\alpha3}(x,y,z) & = C_{\alpha3\beta3}(z)\varphi'_{\beta\mu}(z)\gamma^0_{\mu3}(x,y) \label{eq:transverse_stresses2}\\ 
    \sigma_{33}(x,y,z) & = 0
  \end{empheq}
  \end{subequations}
\fi%
\makeatother%
where $Q_{\alpha\beta\gamma\delta}(z)$ are the generalized plane stress stiffnesses and $C_{\alpha3\beta3}(z)$ are the components of Hooke's tensor corresponding to the transverse shear stiffnesses.
\subsection{Static laminate behavior}
The model requires introduction of generalized forces:
\begin{subequations}\label{eq:generalized_forces1}
\begin{empheq}[left=\empheqlbrace]{align}
  \big\{N_{\alpha\beta},M_{\alpha\beta},P_{\gamma\beta}\big\}&=\int^{h/2}_{-h/2} \big\{1,z,\varphi_{\alpha\gamma}(z)\big\} \sigma_{\alpha\beta}(z) \text{d}z \\
  Q_{\beta}&=\int^{h/2}_{-h/2} \varphi'_{\alpha\beta}(z) \sigma_{\alpha3}(z) \text{d}z
\end{empheq}
\end{subequations}
They are then set, by type, into vectors:
\begin{equation}\label{eq:generalized_forces}
  \mathbf{N}=
  \begin{Bmatrix}
    N_{11} \\
		N_{22} \\
		N_{12}
  \end{Bmatrix}
  ,\ 
  \mathbf{M}=
  \begin{Bmatrix}
    M_{11} \\
		M_{22} \\
	  M_{12}
  \end{Bmatrix}
  ,\ 
  \mathbf{P}=
  \begin{Bmatrix}
    P_{11} \\
		P_{22} \\
		P_{12} \\
	  P_{21}
  \end{Bmatrix}
  ,\ 
  \mathbf{Q}=
  \begin{Bmatrix}
    Q_{1} \\
		Q_{2}
  \end{Bmatrix}
\end{equation} 
and the same is done for the corresponding generalized strains:
\begin{equation}\label{eq:generalized_strains}
  \boldsymbol{\epsilon}=
  \begin{Bmatrix}
    \epsilon^0_{11} \\
		\epsilon^0_{22} \\
		2 \epsilon^0_{12}
  \end{Bmatrix}
  ,\ 
  \boldsymbol{\kappa}=
  \begin{Bmatrix}
    -w^0_{,11} \\
		-w^0_{,22} \\
	  -2w^0_{,12}
  \end{Bmatrix}
  ,\ 
  \mathbf{\Gamma}=
  \begin{Bmatrix}
    \gamma^0_{13,1} \\
		\gamma^0_{23,2} \\
		\gamma^0_{13,2} \\
	  \gamma^0_{23,1}
  \end{Bmatrix}
  ,\ 
  \boldsymbol{\gamma}=
  \begin{Bmatrix}
    \gamma^0_{13} \\
		\gamma^0_{23}
  \end{Bmatrix}
\end{equation} 
Generalized forces are linked with the generalized strains by the $10\times10$ and $2\times2$ following stiffness matrices:
\begin{equation}\label{eq:behavior}
\begin{Bmatrix}
  \mathbf{N} \\
	\mathbf{M} \\
	\mathbf{P}
\end{Bmatrix}
=
\begin{bmatrix}
  \mathbf{A} & \mathbf{B} & \mathbf{E} \\
	\mathbf{B} & \mathbf{D} & \mathbf{F} \\
	\mathbf{E^T} & \mathbf{F^T} & \mathbf{G}
\end{bmatrix}
\begin{Bmatrix}
  \boldsymbol{\epsilon} \\
	\boldsymbol{\kappa} \\
	\boldsymbol{\Gamma}
\end{Bmatrix}
\quad
\begin{Bmatrix}
  \mathbf{Q}
\end{Bmatrix}
=
\begin{bmatrix}
  \mathbf{H}
\end{bmatrix}
\begin{Bmatrix}
	\boldsymbol{\gamma}
\end{Bmatrix}
\end{equation} 
with the following definitions:
\makeatletter%
\if@twocolumn%
  \begin{subequations}\label{eq:generalized stiffnesses1}
  \begin{empheq}[left=\empheqlbrace]{align} 
      \begin{split}
        &\big\{A_{\alpha\beta\gamma\delta},B_{\alpha\beta\gamma\delta},D_{\alpha\beta\gamma\delta},E_{\alpha\beta\mu\delta},F_{\alpha\beta\mu\delta},G_{\nu\beta\mu\delta}\big\} = \int^{h/2}_{-h/2} 
          Q_{\alpha\beta\gamma\delta}(z) \\
        &\phantom{H_{\alpha3\beta3} =} \cdot\big\{1,z,z^2,\varphi_{\gamma\mu}(z),z\varphi_{\gamma\mu}(z),\varphi_{\alpha\nu}(z)\varphi_{\gamma\mu}(z)\big\}\text{d}z
      \end{split} 
      \\
    &H_{\alpha3\beta3} =\int^{h/2}_{-h/2} \varphi'_{\gamma\alpha}(z) C_{\gamma3\delta3}(z) \varphi'_{\delta\beta}(z) \text{d}z
  \end{empheq}
  \end{subequations}
\else%
  \begin{subequations}\label{eq:generalized stiffnesses1}
  \begin{empheq}[left=\empheqlbrace]{align}
    \{A_{\alpha\beta\gamma\delta},B_{\alpha\beta\gamma\delta},D_{\alpha\beta\gamma\delta},E_{\alpha\beta\mu\delta},F_{\alpha\beta\mu\delta},G_{\nu\beta\mu\delta}\}=&\int^{h/2}_{-h/2} Q_{\alpha\beta\gamma\delta}(z) \{1,z,z^2,\varphi_{\gamma\mu}(z),z\varphi_{\gamma\mu}(z), \\ \nonumber &  \varphi_{\alpha\nu}(z)\varphi_{\gamma\mu}(z)\}\text{d}z \\
    H_{\alpha3\beta3}=&\int^{h/2}_{-h/2} \varphi'_{\gamma\alpha}(z) C_{\gamma3\delta3}(z) \varphi'_{\delta\beta}(z) \text{d}z
  \end{empheq}
  \end{subequations}
\fi%
\makeatother%
\subsection{Laminate equations of motion}
\label{sec:LaminateEquationsOfMotion}
Weighted integration of equilibrium equations leads to
\begin{subequations}\label{eq:equilibrium3}
\begin{empheq}[left=\empheqlbrace]{align}
  &N_{\alpha\beta,\beta}=R \ddot{u}^0_{\alpha} - S \ddot{w}^0_{,\alpha}+U_{\alpha\beta}\ddot{\gamma}^0_{\beta3} \\
  &M_{\alpha\beta,\beta\alpha} + q =R \ddot{w}^0 + S \ddot{u}^0_{\alpha,\alpha} - T \ddot{w}^0_{,\alpha\alpha}+V_{\alpha\beta}\ddot{\gamma}^0_{\beta3,\alpha} \\
  &P_{\alpha\beta,\beta}-Q_{\alpha}=U_{\beta\alpha} \ddot{u}^0_{\beta} - V_{\beta\alpha} \ddot{w}^0_{,\beta}+W_{\alpha\beta}\ddot{\gamma}^0_{\beta3} 
\end{empheq}
\end{subequations} 
where $q=[\sigma_{33}(z)]^{h/2}_{-h/2}$ is the value of the transverse loading (no tangential forces are applied on the top and bottom of the plate), and:
\makeatletter%
\if@twocolumn%
\begin{multline}\label{eq:generalized_mass}
  \big\{R,S,T,U_{\alpha\beta},V_{\alpha\beta},W_{\alpha\beta}\big\} = \int^{h/2}_{-h/2} \rho(z)\big\{1,z,z^2,\varphi_{\alpha\beta}(z),\varphi_{\alpha\beta}(z)z,
  \\
  \varphi_{\mu\alpha}(z)\varphi_{\mu\beta}(z)\big\} \text{d}z
\end{multline} 
\else%
\begin{equation}\label{eq:generalized_mass}
  \big\{R,S,T,U_{\alpha\beta},V_{\alpha\beta},W_{\alpha\beta}\big\} = \int^{h/2}_{-h/2} \rho(z)\big\{1,z,z^2,\varphi_{\alpha\beta}(z),\varphi_{\alpha\beta}(z)z,\varphi_{\mu\alpha}(z)\varphi_{\mu\beta}(z)\big\} \text{d}z
\end{equation} 
\fi%
\makeatother%
\section{Warping functions issued from transverse shear stress analysis}\label{sec:WF}
We introduce two different ways to obtain WF from transverse shear stress analysis: a first set of WF is issued from an analytical solution and a second set is issued from an iterative process using the integration of equilibrium equations. First, we shall examine a way to link the WF to the transverse shear stresses.
\subsection{From transverse shear stresses to WF}
\label{sec:FromTrSSToWF}
Considering equation~\eqref{eq:transverse_stresses2}, we see that the $\varphi'_{\alpha\beta}$ are directly linked to the $\sigma_{\alpha3}$. Introducing the transverse shear stresses $\sigma^0_{\delta 3}(x,y)$ at $z=0$ into this equation leads to
\begin{equation}
  \sigma_{\alpha3}(x,y,z) = 4C_{\alpha3\beta3}(z) \varphi'_{\beta\gamma}(z) S_{\gamma3\delta3}(0) \sigma^0_{\delta 3}(x,y)
\end{equation}
where $S_{\gamma3\delta3}$ are components of the compliance tensor. 
\par
This can be written
\begin{equation}\label{eq:psiprime}
  \sigma_{\alpha3}(x,y,z) = \Psi'_{\alpha\beta}(z)\sigma^0_{\beta 3}(x,y)
\end{equation}
where:
\begin{equation}\label{eq:linkphipsi}
  \Psi'_{\alpha\beta}(z) =4 C_{\alpha3\delta3}(z) \varphi'_{\delta \gamma}(z) S_{\gamma3\beta3}(0)
\end{equation}
The $\Psi'_{\alpha\beta}(z)$ cannot be directly issued from equation~\eqref{eq:psiprime} because there are four functions to determine from the variations of two transverse shear stresses, leading to infinitely many solutions. The main idea is to consider a simply supported plate subjected to a double-sine load, and to issue the four functions $\Psi'_{\alpha\beta}(z)$ from the variations of the transverse shear stresses at two separate points of the plate. Since the deformation of the plate is of the form~\eqref{eq:deplacementEE}, the transverse shear stresses at the reference plane are of the form:
\begin{subequations}
\begin{empheq}[left=\empheqlbrace]{align}
		\sigma^0_{13}(x,y) &= s_{13} \cos (\xi x) \sin (\eta y)  + \overline{s}_{13} \sin (\xi x) \cos (\eta y)  \\
		\sigma^0_{23}(x,y) &= s_{23} \sin (\xi x) \cos (\eta y)  + \overline{s}_{23} \cos (\xi x) \sin (\eta y) 
\end{empheq}
\end{subequations} 
These shear stresses are evaluated at points $A(a/2,0)$ and $B(0,b/2)$:
\begin{itemize}
	\item[--] $x=a/2$ and $y=0$ implies $\sigma^0_{13}(A) = \overline{s}_{13}$ and $\sigma^0_{23}(A) = s_{23}$,
	\item[--] $x=0$ and $y=b/2$ implies $\sigma^0_{13}(B) = s_{13}$ and $\sigma^0_{23}(B) = \overline{s}_{23}$.
\end{itemize}
Setting these local values into formula~\eqref{eq:psiprime} leads to the following system:
\begin{equation}
\left[
\begin{matrix}
s_{13} & 0 & \overline{s}_{23} & 0 \\
0 & s_{23} & 0 & \overline{s}_{13} \\
\overline{s}_{13} & 0  & s_{23} & 0 \\
0 & \overline{s}_{23} & 0 &s_{13} \\
\end{matrix}
\right]
\left\{
\begin{matrix}
\Psi'_{11} \\
\Psi'_{22} \\
\Psi'_{12} \\
\Psi'_{21}
\end{matrix}
\right\}
=
\left\{
\begin{matrix}
\sigma_{13}^0(B) \\
\sigma_{23}^0(A) \\
\sigma_{13}^0(A) \\
\sigma_{23}^0(B)
\end{matrix}
\right\}
\end{equation}
The $\Psi'_{\alpha\beta}(z)$ are obtained from the resolution of this system; $\varphi'_{\alpha\beta}(z)$ are then obtained using the reciprocal of equation~\eqref{eq:linkphipsi}:
\begin{equation}
  \varphi'_{\alpha\beta}(z)=4 S_{\alpha3\delta3}(z) \Psi'_{\delta \gamma}(z) C_{\gamma3\beta3}(0)
\end{equation}
Then, integrating the $\varphi'_{\alpha\beta}(z)$ so that $\varphi_{\alpha\beta}(0)=0$ gives the four WF $\varphi_{\alpha\beta}(z)$.
\subsection{WF issued from exact 3D solutions}\label{sec:tdwf}
Exact 3D solutions of simply supported plates subjected to a double-sine load are known for cross-ply and antisymmetric angle-ply laminates since the works of Pagano~\cite{Pagano1970} and Noor~\cite{Noor1990}. Further works have shown that they can be obtained by several ways. Solution for general lamination have been proposed recently in Ref.~\cite{Loredo2014}. The corresponding plate problem is solved with the appropriate method, and the transverse shear stresses are computed at points A and B. Then, the procedure described in the previous section is applied.
\subsection{WF issued from iterative equilibrium equation integration}\label{sec:itwf}
Since transverse shear stresses can be obtained from the equilibrium equations, it is also possible to get the warping functions following an iterative process\footnote{This iterative process, although based on a simpler formulation, has been proposed in the 1989 unpublished reference~\cite{Loredo1989}}. The process, described in the algorithm~\ref{alg:1}, starts with any known WF, says Reddy's $z-4/3(z^3/h^2)$ formula for example, and, at each iteration, the model is updated with WF issued from the transverse stresses of the previous iteration. The procedure is also based on the simply supported plate bending problem with a double-sine loading. Let us establish the needed formulas, starting from the equilibrium conditions within a solid, without body forces:
\begin{subequations}\label{eq:equilibrium0}
\begin{empheq}[left=\empheqlbrace]{align}
  \sigma_{\alpha\beta,\beta}+\sigma_{\alpha3,3}=\rho \ddot{u}_{\alpha} \label{eq:equilibrium0_membrane} \\
  \sigma_{\alpha3,\alpha}+\sigma_{33,3}=\rho \ddot{u}_{3}
\end{empheq}
\end{subequations} 
The transverse shear stresses, for the static case, are therefore computed using:
\makeatletter
\if@twocolumn
  \begin{multline}\label{eq:equilibrium1}
  \sigma_{\alpha3}(z)=-\int_{-h/2}^z\sigma_{\alpha\beta,\beta}(z)\text{d}z = -\int_{-h/2}^z Q_{\alpha\beta\gamma\delta}(z)\Big[u_{\gamma,\delta\beta}^0(x,y) \\
     - z w^0_{,\gamma\delta\beta}(x,y) +  \varphi_{\gamma\mu}(z)\gamma^0_{\mu3,\delta\beta}(x,y)\Big] \text{d}z
  \end{multline}
\else
  \begin{align}\label{eq:equilibrium1}\nonumber
  \sigma_{\alpha3}(z)&=-\int_{-h/2}^z\sigma_{\alpha\beta,\beta}(z)\text{d}z \\
                     &=-\int_{-h/2}^z Q_{\alpha\beta\gamma\delta}(z)\left(u_{\gamma,\delta\beta}^0(x,y) - z w^0_{,\gamma\delta\beta}(x,y) + \varphi_{\gamma\mu}(z)\gamma^0_{\mu3,\delta\beta}(x,y)\right) \text{d}z
  \end{align}
\fi
\makeatother
Then, the spatial derivatives of the generalized displacements are eliminated accounting to the specific nature of the chosen functions in formulas~\eqref{eq:deplacementEE}.
\par
\vskip 0.5em
\begin{algorithm}
  Set i=0, $w_0=0$\;
  Take Reddy's formula as WF\;
  \Repeat{$\left|\frac{w_i-w_{i-1}}{w_i}\right| > \epsilon$}{
    i=i+1\;
    Compute stiffness matrices\;
    Solve the displacements with Navier's method\;
    Assign the deflection to $w_i$\;
    Compute transverse shear stresses using equation~\eqref{eq:equilibrium1}\;
    Compute new WF using the procedure of section~\ref{sec:FromTrSSToWF}\;
  }
\caption{Obtaining warping functions from equilibrium equations\label{alg:1}}
\end{algorithm}
\subsection{Discussion}
The two ways to obtain WF from transverse shear stresses described in the above sections are based on the study of a simply supported plate subjected to a double-sine load. As one shall see in the following, the two methods give very similar results. This proves that the model, which is strongly implicated in the iterative process, is able to fit the transverse shear stresses of the 3D solution with good agreement. Of course, there is no guarantee at this time that these WF will be the best candidates if another plate problem is studied, with different boundary conditions and/or different loading. It is precisely the reason why the iterative process is interesting because it might be adapted to a local strategy.

\section{Solving method by a Navier-like procedure}\label{sec:Navier}
A Navier-like procedure is implemented to solve both static and dynamic problems for a simply supported plate. For the static case, in order to respect the simply supported boundary condition for laminates which are not of cross-ply nor anti-symmetrical angle-ply types, a specific loading is applied. This is done with the help of a Lagrange multiplier, as explained below. The dynamic study is restricted to the search of the natural frequencies.
\par
The Fourier series is limited to one term, hence the generalized displacement field is
\begin{equation}
	\label{eq:deplacementEE}
	\left\{
	\begin{array}{c}
		u_1 \\
		u_2 \\
		w \\
		\gamma_{13} \\
		\gamma_{23}
	\end{array}
	\right\}
	=
  \left\{
	\begin{array}{clllll}
		u^{mn}_1            & \cos (\xi x) & \sin (\eta y)  &+	 \overline{u}^{mn}_1    & \sin (\xi x) & \cos (\eta y)  \\
		u^{mn}_2            & \sin (\xi x) & \cos (\eta y)  &+	 \overline{u}^{mn}_2    & \cos (\xi x) & \sin (\eta y)  \\
		w^{mn}              & \sin (\xi x) & \sin (\eta y)  &+	 \overline{w}^{mn}      & \cos (\xi x) & \cos (\eta y)  \\
		\gamma^{mn}_{13}    & \cos (\xi x) & \sin (\eta y)  &+	 \overline{\gamma}^{mn}_{13}    & \sin (\xi x) & \cos (\eta y)  \\
		\gamma^{mn}_{23}    & \sin (\xi x) & \cos (\eta y)  &+	 \overline{\gamma}^{mn}_{23}    & \cos (\xi x) & \sin (\eta y)  \\
	\end{array}
	\right\}
\end{equation}
with
\begin{equation}\nonumber
\xi = \frac{m \pi}{a} \text{ and } \eta = \frac{n \pi}{b}
\end{equation}
where $a$ and $b$ are the length of the sides of the plate, $m$ and $n$ are wavenumbers, set to $1$ for static analysis or to arbitrary values for the dynamic study of the corresponding mode. Then for a given $m$ and $n$, the motion equations of section~\ref{sec:LaminateEquationsOfMotion} give a stiffness and a mass matrix, respectively $\mymat{K}$ and $\mymat{M}$, related to the vector $\myvect{U}=\{u_1^{mn},u_2^{mn}\dots\overline{\gamma}^{mn}_{23}\}$. The static case is treated by solving the linear system $\mymat{K}\myvect{U}=\myvect{F}$, where $\myvect{F}$ is a force vector containing $q^{mn}$ for its third component (generally set to one). Solving the dynamic case consists in searching the generalized eigenvalues for matrices $\mymat{K}$ and $\mymat{M}$.
For cross-ply and antisymmetric angle ply, $w$ respects the simply supported conditions, i. e. $\overline{w}_{mn}=0$. For the general laminates, the deflection under a double-sine loading gives a $\overline{w}_{mn} \neq 0$. Since we choose to keep simply supported boundary conditions, $\overline{w}_{mn}$ may be set to zero if a bi-cosine term is added to the loading. The amplitude of the bi-cosine term $\overline{q}^{mn}$ is obtained using a Lagrange multiplier. The stiffness matrix is then of size $11 \times 11$.
\begin{equation}
\left[ 
\begin {array}{c|c}
  \mathbf{K}&\mathbf{C} \\ \noalign{\medskip} \hline \noalign{\medskip}
 	\mathbf{C}^T& 0
 \end{array} 
 \right] 
\left\{
\begin{array}{c}
\mathbf{U} \\  \noalign{\medskip} \hline \noalign{\medskip} \overline{q}_{mn}
\end{array}
\right\}
=
\left\{
\begin{array}{c}
\mathbf{F} \\ \noalign{\medskip} \hline \noalign{\medskip} 0
\end{array}
\right\}
\label{eq:syslin}
\end{equation}
with $\myvect{C}$ being a vector with a one on it's eighth component. For general laminates, this augmented system remains regular but no longer positive definite, hence the solving procedure must be chosen adequately. This is not detrimental because the system is of very small size. For the dynamic case, the matrix $\mymat{M}$ is augmented with a line and a column of zeros so it becomes a $11 \times 11$ matrix.
Detailed formulation of matrices $\mymat{K}$ and $\mymat{M}$ are given in~\ref{sec:detailedmatrices}. Note that it is also possible to keep the loading of the form $q(x,y)=q^{mn}\sin(\xi x)\sin (\eta y)$, then the simply supported boundary condition is no longer respected for general laminates.
\section{Reference models}
\label{sec:RefModels}
The model using WF issued from transverse shear stresses of analytical solutions, presented in section~\ref{sec:tdwf}, will be denoted \tdwf{}. The model using WF obtained by the iterative process, presented in section~\ref{sec:itwf}, will be denoted \itwf{}. The results obtained with \tdwf{} and \itwf{} models are compared to those obtained with different models issued from the literature and with exact analytical solutions. These reference models are presented below.
\subsection{Exact analytical solutions}\label{sec:ExaAnaSolu}
\subsubsection{The Exa solution}\label{sec:Exa}
Each studied case is solved by a state-space algorithm described in reference~\cite{Loredo2014}. The algorithm is a generalization for general lamination sequences of existing algorithms for cross-ply and antisymmetric angle-ply lamination sequences. In the reference~\cite{Loredo2014}, the algorithm has been tested: exact solutions of previous works have been reproduced, and for general lamination sequences, the solution has been confronted with success to an accurate finite element computation. In this work, the algorithm is used to obtain deflections, stresses and natural frequencies taken as reference for comparisons, but it is also used to create a set of WF for the \tdwf{} model, as explained in section~\ref{sec:WF}. The corresponding solution is denoted Exa in the following text and in tables. For all the studied configurations, the loading is divided into two equal parts which are applied to the top and bottom faces.
\subsubsection{The Exa$^2$ solution}\label{sec:Exa2}
All models involved in this study consider the generalized plane stress assumption which leads to the use of the reduced stiffnesses $Q_{\alpha\beta\gamma\delta}=C_{\alpha\beta\gamma\delta}-C_{\alpha\beta33}C_{33\gamma\delta}/C_{3333}$. Further, these models do not consider the normal strain $\varepsilon_{33}$ in their formulation. It is well known that these assumptions are correct when the length-to-thickness ratio is high and when layers are made of similar materials. This hypothesis is no longer valid for low length-to-thickness ratio (say 2 and 4) and when materials with very different behavior are used, which is typical of sandwiches. Hence, it is not possible to draw a conclusion from a comparison of models with an analytical solution they cannot approximate. For this reason, another exact solution is computed for a virtual laminate which has modified stiffness values in order to, firstly, fit the generalized plane stress assumption and secondly, force $\varepsilon_{33}\approx0$. It is done by setting $E_3=10^{10} E_2$ and $\nu_{13}=\nu_{23}=0$ for each material. This is equivalent to the replacement of the $C_{\alpha\beta\gamma\delta}$ by the $Q_{\alpha\beta\gamma\delta}$ and forces $\varepsilon_{33}$ to take very small values. This exact solution, designated by the Exa$^2$ symbol, is computed with the same algorithm than the Exa solution. For this case too, the loading is divided into two equal parts which are applied to the top and bottom faces. 
\subsection{Other models}\label{sec:OtherModels}
Some more or less classical models have been chosen for comparison matters. They offer the advantage to be easily simulated by the present model when appropriate sets of WF are selected:
\begin{itemize}
	\item First-order Shear Deformation Theory (denoted FoSDT in tables and figures): Often called Mindlin plate theory, it can be formulated setting in the generic model the following warping functions,
	\begin{equation}
		\varphi_{\alpha\beta}(z)=\delta^K_{\alpha\beta} z
	\end{equation}
	where $\delta^K_{\alpha\beta}$ is Kronecker's delta. Note that this theory is generally used with shear correction factors, often the $5/6$ factor which corresponds to an homogeneous plate. As there are several ways to compute these correction factors in the general case, we chose in this study not to use them. This is of course a serious penalty for this model.
	\item Third-order Shear Deformation Theory (denoted ToSDT): often called Reddy's third order theory, verifies that transverse shear stresses are null at the top and bottom faces of the plate. It is simulated using the following warping functions:
	\begin{equation}
		\varphi_{\alpha \beta}(z) = \delta^K_{\alpha \beta} \left(z - \frac{4}{3}\frac{z^3}{h^2}\right)
	\end{equation}
	\item First-order Zig-Zag model with 4 WF (denoted \fozz{}): This model verifies the continuity of transverse shear stresses at the layers' interfaces. It was first presented by Sun \& Whitney~\cite{Sun1973} for cross-ply laminates and generalized for general lamination sequences by Woodcock~\cite{Woodcock2008}. It is possible to formulate this model with the following WF as shown in~\cite{Loredo2013}:
	\begin{equation}
		\varphi_{\alpha\beta}(z)=4 C_{\gamma3\beta3}(0) \int_{-h/2}^{z} S_{\alpha3\gamma3}(\zeta) \text{d}\zeta
	\end{equation}
	\item Third-order Zig-Zag model with 4 WF: (denoted \tozz{}): This formulation, presented by Cho\footnote{In reference~\cite{Carrera2002}, Carrera wrote that this model was a re-discovery of previous works, and called the model the Ambartsumyan--Whitney--Rath--Das theory.}~\cite{Cho1993, Kim2007} consists in superimposing a cubic displacement field, which permits the transverse shear stresses to be null at the top and bottom faces of the laminate, to a zig-zag displacement field issued from the continuity of the transverse shear stresses at layer interfaces. The corresponding WF, which are polynomials of third order in $z$, are not detailed. Indeed, as their computation involves the resolution of a system of equations, it is difficult to give here an explicit form. Note also that for coupled laminates, an extension of this model~\cite{Kim2006} has been proposed. This extension has not been implemented in this study, it could have given different results for the studied $[-15/15]$ case (which is the only coupled laminate considered in this study).
\end{itemize}
As these models can be implemented by means of WF, the solving process described in section~\ref{sec:Navier} can be applied to all models.

\section{Numerical results}
\label{sec:nresults}
This section proposes the study of five configurations including four laminates and a sandwich plate. Only two materials are involved: an orthotropic composite material used in all laminates and an honeycomb-type material used for the core of the sandwich panel. All the material properties are given in table~\ref{tab:matprop}.
\par
\begin{table*}[!htb]
	\centering
	\small
	\begin{tabular}{@{}lcccccccccc@{}}
	                    &   $E_1$   &       $E_2$     &     $E_3$   &  $G_{23}$  &  $G_{13}$  &  $G_{12}$  & $\nu_{23}$ & $\nu_{13}$ & $\nu_{12}$ & $\rho$ \\	\hline
    Composite ply (p) & $25E^p_2$ &      $10^6$     &     $E^p_2$ & $0.2E^p_2$ & $0.5E^p_2$ & $0.5E^p_2$ &   $0.25$   &   $0.25$   &   $0.25$   & $1500$ \\
		Core material (c) &  $E^c_2$  & $4 \times 10^4$ & $12.5E^c_2$ & $1.5E^c_2$ & $1.5E^c_2$ & $1.5E^c_2$ &   $0.25$   &   $0.25$   &   $0.25$   &  $100$ 
	\end{tabular}
	\caption{Material properties}
	 \label{tab:matprop}
\end{table*}
\par
Three nondimensionalized quantities are considered and compared to those issued from analytical solutions:
\begin{itemize}
	\item[--] Deflections $w$ are nondimensionalized using equation
	\begin{equation}
	w^{*}=100\frac{E_2^{\text{ref}} h^3}{(-q)a^4}w 
	\label{eq:nondimd}
\end{equation}
	\item[--] First natural frequencies are nondimensionalized using equation
  \begin{equation}
	  \omega^{*}=\frac{a^2}{h}\sqrt{\frac{\rho^{\text{ref}}}{E_2^{\text{ref}}}}\omega 
	  \label{eq:nondimf}
  \end{equation}
	\item[--] Shear stresses are nondimensionalized using equation
  \begin{equation}
	  \sigma_{\alpha3}^{*}=10\frac{h}{(-q)a}\sigma_{\alpha3}
	  \label{eq:nondims}
  \end{equation}
\end{itemize}
where $E_2^{\text{ref}}$ and $\rho^{\text{ref}}$ are taken as values of the core material for the sandwich and as values of the composite ply for laminates.
\subsection{Rectangular $[0/90/0]$ cross-ply composite plate}\label{sec:p0p90p0_x3}
This plate is made of three composite plies of equal thickness, with a $[0/90/0]$ stacking sequence and $b=3a$. Results are presented in table~\ref{tab:p0p90p0}. Note that Reddy's model (ToSDT) -- with relatively simple WF -- gives quite good results for this laminate. Cho's model (\tozz{}) -- with more sophisticated WF -- gives better values than the previous, except for the $a/h=2$ length-to-thickness ratio. Results also show that the two proposed models, \tdwf{} and \itwf{}, give a very satisfying accuracy for all length-to-thickness ratios. Compared to other models with reference to the exact solution, values for the deflection are among the best results, values for the transverse stresses at points A and B, and for the first natural frequency, are the best. However, we can note a weakness for the prediction of $\sigma_{23}(A)$, which is probably due to a sandwich-like behavior of this laminate according to the $y$ direction. The \fozz{} model, which is accurate with sandwiches, tends to confirm this point. Both proposed models, even though warping functions are generated with two very different methods, give almost identical results. Compared to the Exa$^2$ plane stress exact solution, \tdwf{} and \itwf{} models give the best results. The model \itwf{} gives, for this problem, the same result as the exact solution. This point will be discussed later on section~\ref{sec:discussion}.
\par
Figure~\ref{fig:p0p90p0_x3_phi} shows the corresponding WF for $a/h=4$, for all plate models except the FoSDT one. Figure~\ref{fig:p0p90p0_x3_sigma} presents the variations of the transverse shear stresses obtained by integration of the equilibrium equations for all models, compared to the exact solution, in the $a/h=4$ case. 
\iftoggle{submission}{}{\tikzsetnextfilename{p0p90p0_x3_phi}}
\begin{SmartFigure}[!htb]%
  \centering
  \iftoggle{submission}{
    \includegraphics{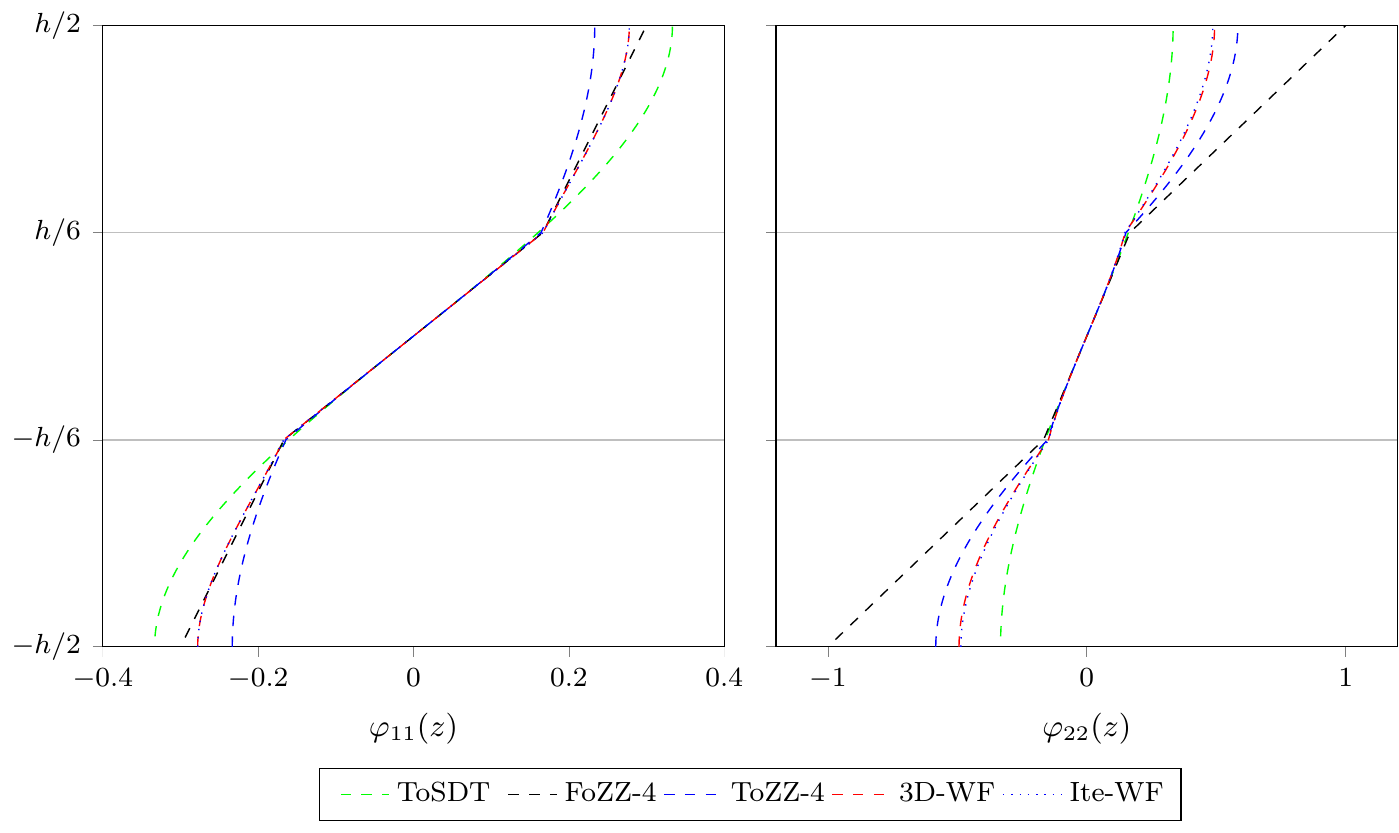}
  }{
    \input{Figures/p0p90p0_x3_phi.tex}%
  }
  \caption{Transverse shear WF of the rectangular $[0/90/0]$ composite plate with $a/h=4$ for each model.}%
  \label{fig:p0p90p0_x3_phi}%
\end{SmartFigure}
\iftoggle{submission}{}{\tikzsetnextfilename{p0p90p0_x3_sigma}}
\begin{SmartFigure}[!htb]%
  \centering
  \iftoggle{submission}{
    \includegraphics{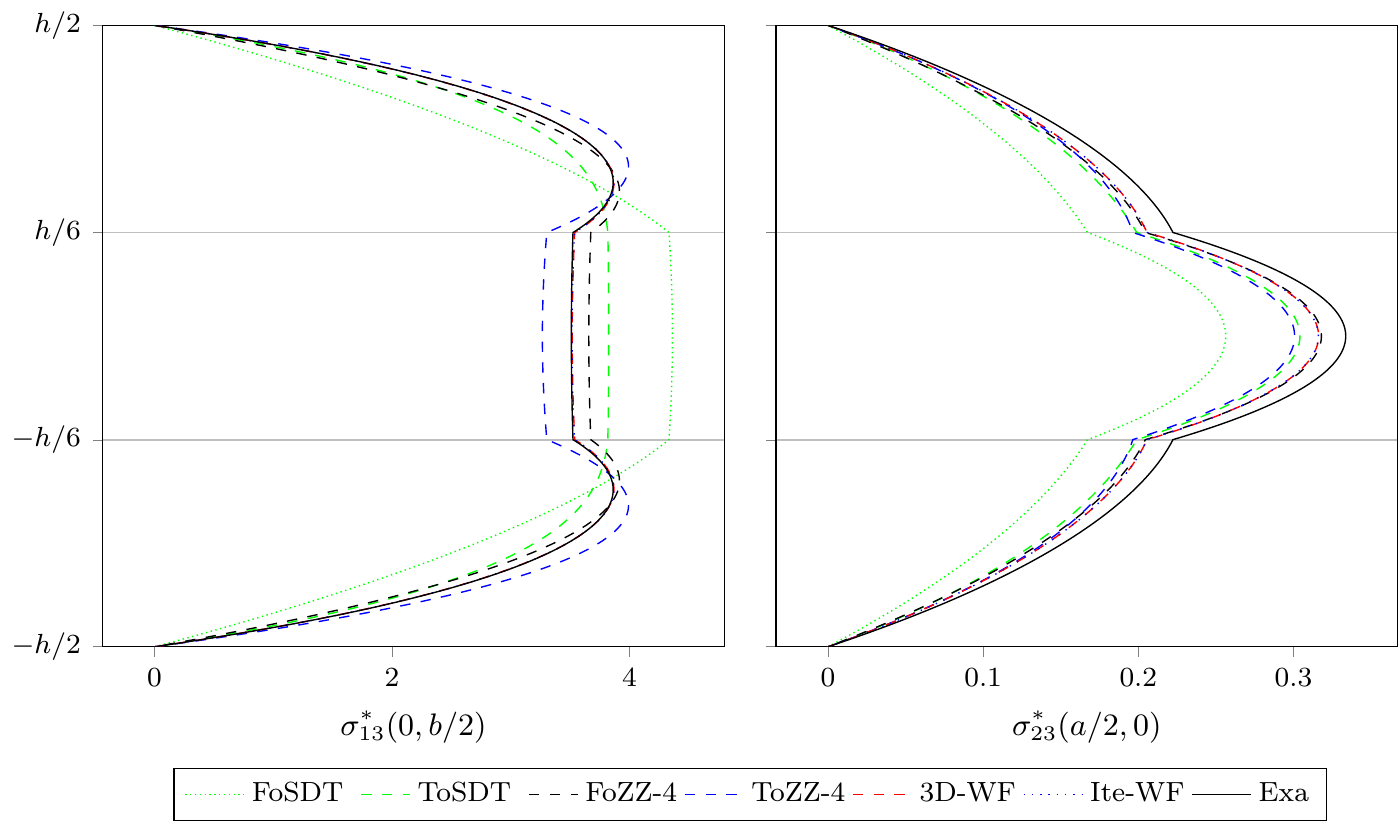}
  }{
    \input{Figures/p0p90p0_x3_sigma.tex}%
  }
  \caption{Nondimensionalized transverse shear stresses of the rectangular $[0/90/0]$ composite plate with $a/h=4$ for each model.}%
  \label{fig:p0p90p0_x3_sigma}%
\end{SmartFigure}
\begin{table*}[!htb]%
	\centering
	\footnotesize
	\begin{tabular}{@{~~}r@{~~}l|l@{~~}r@{~~}r|l@{~~}r@{~~}r|l@{~~}r@{~~}r|l@{~~}r@{~~}r@{~~}}
    a/h & Model & $w^{*}$ & \%~~ & \%~~ & $\sigma_{13}(B)$ & \%~~ & \%~~ & $\sigma_{23}(A)$ & \%~~ & \%~~ & $\omega^{*}$ & \%~~ & \%~~ \\ 

    \hline
    2 & FoSDT  & $6.6164$ & $-18.98$ & $-21.69$ & $4.2978$ & $+67.17$ & $+64.28$ & $0.54919$ & $-17.76$ & $-13.75$ & $3.8633$ & $+12.97$ & $+13.01$ \\ 
  & ToSDT & $7.8944$ &  $-3.32$ &  $-6.56$ & $2.4649$ &  $-4.12$ &  $-5.78$ & $0.59477$ & $-10.93$ &  $-6.59$ & $3.5335$ &  $+3.33$ &  $+3.36$ \\ 
  & \fozz  & $7.8130$ &  $-4.32$ &  $-7.52$ & $2.0026$ & $-22.11$ & $-23.45$ & $0.66040$ &  $-1.11$ &  $+3.72$ & $3.5485$ &  $+3.76$ &  $+3.80$ \\ 
  & \tozz & $6.4960$ & $-20.45$ & $-23.11$ & $1.4032$ & $-45.42$ & $-46.36$ & $0.47682$ & $-28.60$ & $-25.11$ & $3.8845$ & $+13.59$ & $+13.63$ \\ 
  & \tdwf  & $8.4451$ &  $+3.42$ &  $-0.04$ & $2.6931$ &  $+4.76$ &  $+2.94$ & $0.64111$ &  $-3.99$ &  $+0.69$ & $3.4196$ &  $-0.00$ &  $+0.03$ \\ 
  & \itwf & $8.4487$ &  $+3.46$ &  $-0.00$ & $2.6163$ &  $+1.76$ &  $+0.00$ & $0.63670$ &  $-4.65$ &  $+0.00$ & $3.4186$ &  $-0.03$ &  $+0.00$ \\ 
  & Exa$^2$ & $8.4488$ &  & \emph{ref.}  & $2.6162$ &  & \emph{ref.}      & $0.63670$ &  & \emph{ref.}      & $3.4185$ &  & \emph{ref.} \\ 
  & Exa & $8.1659$ & \emph{ref.} &  & $2.5709$ & \emph{ref.} &  & $0.66779$ & \emph{ref.} &  & $3.4197$ & \emph{ref.} &  \\ 

    \hline
    4 & FoSDT  & $2.0547$ & $-27.17$ & $-27.80$ & $4.3625$ & $+24.26$ & $+24.05$ & $0.25631$ & $-23.18$ & $-18.89$ & $6.9503$ & $+17.60$ & $+17.64$ \\ 
  & ToSDT & $2.6411$ & $-6.38$ & $-7.20$ & $3.8253$ & $+8.96$ & $+8.78$ & $0.30414$ & $-8.84$ & $-3.75$ & $6.1331$ & $+3.77$ & $+3.80$ \\ 
  & \fozz  & $2.7172$ & $-3.68$ & $-4.52$ & $3.6580$ & $+4.19$ & $+4.02$ & $0.31797$ & $-4.70$ & $+0.62$ & $6.0461$ & $+2.30$ & $+2.33$ \\ 
  & \tozz & $2.7331$ & $-3.12$ & $-3.96$ & $3.2669$ & $-6.95$ & $-7.10$ & $0.30071$ & $-9.87$ & $-4.84$ & $6.0265$ & $+1.97$ & $+2.00$ \\ 
  & \tdwf  & $2.8459$ & $+0.88$ & $-0.00$ & $3.5182$ & $+0.21$ & $+0.04$ & $0.31616$ & $-5.24$ & $+0.05$ & $5.9084$ & $-0.03$ & $+0.00$ \\ 
  & \itwf & $2.8459$ & $+0.88$ & $-0.00$ & $3.5167$ & $+0.17$ & $+0.00$ & $0.31600$ & $-5.29$ & $+0.00$ & $5.9084$ & $-0.03$ & $+0.00$ \\ 
  & Exa$^2$ & $2.8459$ &  & \emph{ref.} & $3.5167$ &  & \emph{ref.} & $0.31600$ &  & \emph{ref.} & $5.9083$ &  & \emph{ref.} \\ 
  & Exa & $2.8211$ & \emph{ref.} &  & $3.5108$ & \emph{ref.} &  & $0.33365$ & \emph{ref.} &  & $5.9100$ & \emph{ref.} &  \\ 

    \hline
    10 & FoSDT  & $0.75314$ & $-18.04$ & $-18.19$ & $4.3895$ & $+4.48$ & $+4.48$ & $0.13418$ & $-11.94$ & $-9.77$ & $11.497$ & $+10.47$ & $+10.49$ \\ 
   & ToSDT & $0.86219$ & $-6.17$ & $-6.34$ & $4.2988$ & $+2.32$ & $+2.32$ & $0.14468$ & $-5.05$ & $-2.71$ & $10.750$ & $+3.29$ & $+3.31$ \\ 
   & \fozz  & $0.88102$ & $-4.12$ & $-4.30$ & $4.2687$ & $+1.60$ & $+1.61$ & $0.14659$ & $-3.80$ & $-1.42$ & $10.635$ & $+2.19$ & $+2.21$ \\ 
   & \tozz & $0.91831$ & $-0.07$ & $-0.25$ & $4.1841$ & $-0.41$ & $-0.41$ & $0.14813$ & $-2.79$ & $-0.39$ & $10.418$ & $+0.10$ & $+0.12$ \\ 
   & \tdwf  & $0.92059$ & $+0.18$ & $-0.00$ & $4.2011$ & $-0.01$ & $-0.00$ & $0.14871$ & $-2.41$ & $+0.00$ & $10.405$ & $-0.02$ & $+0.00$ \\ 
   & \itwf & $0.92059$ & $+0.18$ & $-0.00$ & $4.2012$ & $-0.00$ & $+0.00$ & $0.14871$ & $-2.41$ & $+0.00$ & $10.405$ & $-0.02$ & $+0.00$ \\ 
   & Exa$^2$ & $0.92059$ &  & \emph{ref.} & $4.2012$ &  & \emph{ref.} & $0.14870$ &  & \emph{ref.} & $10.405$ &  & \emph{ref.} \\ 
   & Exa & $0.91891$ & \emph{ref.} &  & $4.2014$ & \emph{ref.} &  & $0.15237$ & \emph{ref.} &  & $10.408$ & \emph{ref.} &  \\ 

    \hline
    100 & FoSDT  & $0.50588$ & $-0.35$ & $-0.35$ & $4.3952$ & $+0.05$ & $+0.05$ & $0.10816$ & $-0.18$ & $-0.15$ & $14.059$ & $+0.18$ & $+0.18$ \\ 
    & ToSDT & $0.50700$ & $-0.13$ & $-0.13$ & $4.3943$ & $+0.03$ & $+0.03$ & $0.10827$ & $-0.08$ & $-0.04$ & $14.043$ & $+0.07$ & $+0.07$ \\ 
    & \fozz  & $0.50721$ & $-0.09$ & $-0.09$ & $4.3940$ & $+0.02$ & $+0.02$ & $0.10829$ & $-0.06$ & $-0.03$ & $14.041$ & $+0.05$ & $+0.05$ \\ 
    & \tozz & $0.50766$ & $+0.00$ & $-0.00$ & $4.3931$ & $-0.00$ & $-0.00$ & $0.10832$ & $-0.04$ & $-0.00$ & $14.034$ & $+0.00$ & $+0.00$ \\ 
    & \tdwf  & $0.50767$ & $+0.00$ & $-0.00$ & $4.3932$ & $-0.00$ & $+0.00$ & $0.10832$ & $-0.04$ & $+0.00$ & $14.034$ & $-0.00$ & $+0.00$ \\ 
    & \itwf & $0.50767$ & $+0.00$ & $-0.00$ & $4.3932$ & $-0.00$ & $+0.00$ & $0.10832$ & $-0.04$ & $+0.00$ & $14.034$ & $-0.00$ & $+0.00$ \\ 
    & Exa$^2$ & $0.50767$ &  & \emph{ref.} & $4.3932$ &  & \emph{ref.} & $0.10832$ &  & \emph{ref.} & $14.034$ &  & \emph{ref.} \\ 
    & Exa & $0.50766$ & \emph{ref.} &  & $4.3932$ & \emph{ref.} &  & $0.10836$ & \emph{ref.} &  & $14.034$ & \emph{ref.} &  \\ 

	\end{tabular}
	\normalsize
	\caption{Comparison between the different models for the rectangular $[0/90/0]$ composite plate with a varying length-to-thickness ratio.}
	\label{tab:p0p90p0}
\end{table*}
\subsection{Square $[0/c/0]$ sandwich plate}\label{sec:Sandwich}
In order to study the behavior of a structure exhibiting a high variation of stiffness through the thickness, we propose to study a square sandwich plate with ply thicknesses $h_1=h_3=0.1 h$ and $h_2 = 0.8 h$. The face sheets are made of one ply of unidirectional composite and the core is constituted of a honeycomb-type material. Material properties are presented in table~\ref{tab:matprop}. Results presented in table~\ref{tab:Sandwich} show that, this time, Reddy's model (ToSDT) is not as accurate as in the previous case, although Cho's model (\tozz{}) obtains better values. This is due to the particular nature of sandwich materials which gives typically zig-zag variations for displacements through the thickness and then typically zig-zag WF. Cho's model (\tozz{}) is able to fit this kind of variation although Reddy's (ToSDT) model is not. The Sun~\&~Withney model (\fozz{}) has also been proved to be very efficient for sandwiches, which can be verified in this table. Note that the two proposed models, \tdwf{} and \itwf{}, globally give the best results. Comparison with the Exa$^2$ model is discussed in section~\ref{sec:discussion}.
\par
Figure~\ref{fig:Sandwich_phi} shows the corresponding WF for $a/h=4$, for all plate models except the FoSDT one. Figure~\ref{fig:Sandwich_sigma} presents the variations of the transverse shear stresses obtained by integration of the equilibrium equations for all models, compared to the exact solution, in the $a/h=4$ case.   
\iftoggle{submission}{}{\tikzsetnextfilename{Sandwich_phi}}
\begin{SmartFigure}[!htb]%
  \centering
  \iftoggle{submission}{
    \includegraphics{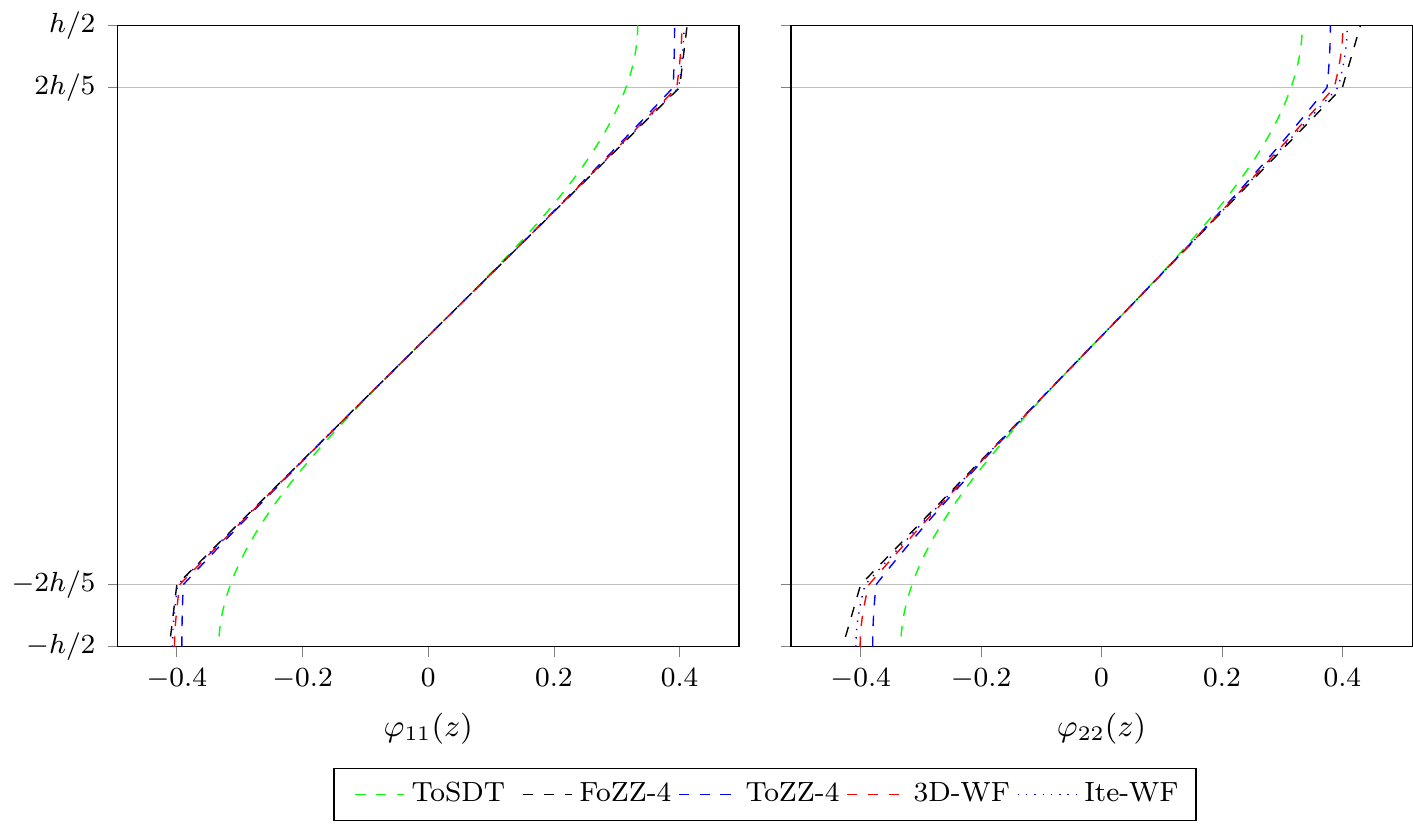}
  }{
    \input{Figures/Sandwich_phi.tex}%
  }
  \caption{Transverse shear WF of the $[0/c/0]$ sandwich plate with $a/h=4$ for each model.}%
  \label{fig:Sandwich_phi}
\end{SmartFigure}
\iftoggle{submission}{}{\tikzsetnextfilename{Sandwich_sigma}}
\begin{SmartFigure}[!htb]%
  \centering
  \iftoggle{submission}{
    \includegraphics{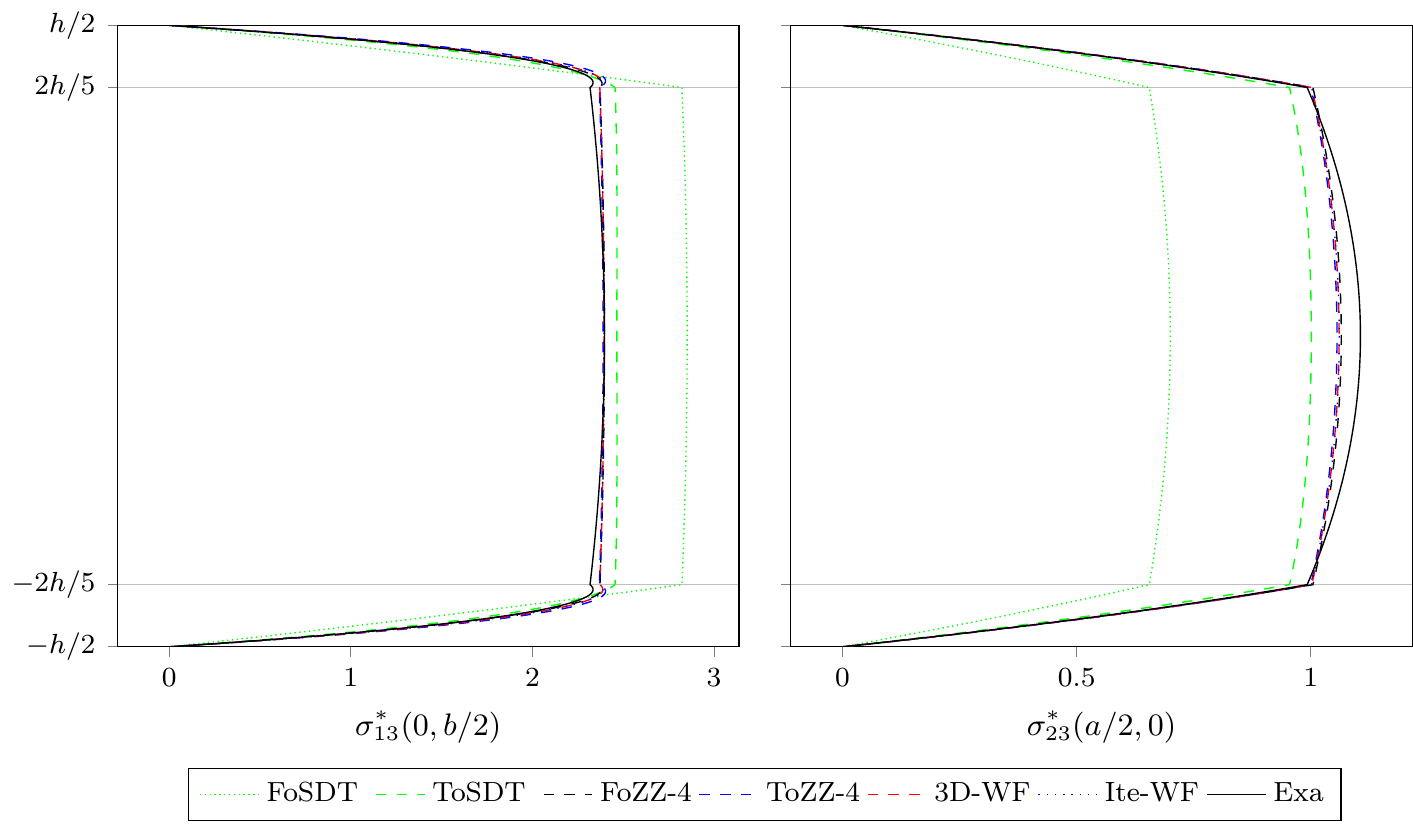}
  }{
    \input{Figures/Sandwich_sigma.tex}%
  }
  \caption{Nondimensionalized transverse shear stresses of the $[0/c/0]$ sandwich plate with $a/h=4$ for each model.}%
  \label{fig:Sandwich_sigma}
\end{SmartFigure}
\begin{table*}[!htb]%
	\centering
	\footnotesize
	\begin{tabular}{@{~~}r@{~~}l|l@{~~}r@{~~}r|l@{~~}r@{~~}r|l@{~~}r@{~~}r|l@{~~}r@{~~}r@{~~}}
    a/h & Model & $w^{*}$ & \%~~ & \%~~ & $\sigma_{13}(B)$ & \%~~ & \%~~ & $\sigma_{23}(A)$ & \%~~ & \%~~ & $\omega^{*}$ & \%~~ & \%~~ \\ 

    \hline
    2 & FoSDT  & $0.50904$ & $-44.74$ & $-44.07$ & $2.5177$ & $+31.01$ & $+35.10$ & $1.0476$ & $-30.40$ & $-24.36$ & $6.9339$ & $+29.88$ & $+31.57$ \\ 
  & ToSDT & $0.85344$ &  $-7.34$ &  $-6.23$ & $1.9150$ &  $-0.35$ &  $+2.76$ & $1.3300$ & $-11.65$ &  $-3.97$ & $5.4307$ &  $+1.72$ & $+3.04$ \\ 
  & \fozz  & $0.90224$ &  $-2.05$ &  $-0.86$ & $1.8459$ &  $-3.95$ &  $-0.95$ & $1.3963$ &  $-7.24$ &  $+0.82$ & $5.2900$ &  $-0.92$ & $+0.37$ \\ 
  & \tozz & $0.88896$ &  $-3.49$ &  $-2.32$ & $1.8320$ &  $-4.67$ &  $-1.70$ & $1.3570$ &  $-9.85$ &  $-2.02$ & $5.3281$ &  $-0.20$ & $+1.10$ \\ 
  & \tdwf  & $0.90894$ &  $-1.32$ &  $-0.13$ & $1.8562$ &  $-3.41$ &  $-0.40$ & $1.3795$ &  $-8.36$ &  $-0.40$ & $5.2730$ &  $-1.23$ & $+0.05$ \\ 
  & \itwf & $0.91010$ &  $-1.19$ &  $-0.00$ & $1.8636$ &  $-3.02$ &  $+0.00$ & $1.3850$ &  $-7.99$ &  $+0.00$ & $5.2703$ &  $-1.28$ & $+0.00$ \\ 
  & Exa$^2$ & $0.91010$ &  & \emph{ref.}  & $1.8636$ &  & \emph{ref.}      & $1.3850$ &  & \emph{ref.}      & $5.2702$ &  & \emph{ref.} \\ 
  & Exa & $0.92108$ & \emph{ref.} &  & $1.9217$ & \emph{ref.} &  & $1.5053$ & \emph{ref.} &  & $5.3389$ & \emph{ref.} &  \\ 

    \hline
    4 & FoSDT  & $0.16645$ & $-45.57$ & $-45.67$ & $2.8525$ & $+18.99$ & $+19.09$ & $0.69992$ & $-36.72$ & $-34.02$ & $12.233$ & $+33.77$ & $+34.67$ \\ 
  & ToSDT & $0.28349$ & $-7.30$ & $-7.47$ & $2.4657$ & $+2.85$ & $+2.94$ & $1.0011$ & $-9.49$ & $-5.63$ & $9.4350$ & $+3.17$ & $+3.86$ \\ 
  & \fozz  & $0.30453$ & $-0.42$ & $-0.60$ & $2.3971$ & $-0.01$ & $+0.08$ & $1.0651$ & $-3.70$ & $+0.41$ & $9.1084$ & $-0.40$ & $+0.27$ \\ 
  & \tozz & $0.30416$ & $-0.54$ & $-0.72$ & $2.3897$ & $-0.32$ & $-0.23$ & $1.0560$ & $-4.52$ & $-0.45$ & $9.1152$ & $-0.33$ & $+0.34$ \\ 
  & \tdwf  & $0.30636$ & $+0.18$ & $-0.01$ & $2.3947$ & $-0.11$ & $-0.02$ & $1.0603$ & $-4.13$ & $-0.04$ & $9.0843$ & $-0.67$ & $+0.00$ \\ 
  & \itwf & $0.30638$ & $+0.19$ & $-0.00$ & $2.3952$ & $-0.09$ & $+0.00$ & $1.0607$ & $-4.09$ & $+0.00$ & $9.0840$ & $-0.67$ & $+0.00$ \\ 
  & Exa$^2$ & $0.30638$ &  & \emph{ref.} & $2.3952$ &  & \emph{ref.} & $1.0607$ &  & \emph{ref.} & $9.0840$ &  & \emph{ref.} \\ 
  & Exa & $0.30581$ & \emph{ref.} &  & $2.3973$ & \emph{ref.} &  & $1.1060$ & \emph{ref.} &  & $9.1452$ & \emph{ref.} &  \\ 

    \hline
    10 & FoSDT  & $0.057970$ & $-34.15$ & $-34.25$ & $3.1506$ & $+5.10$ & $+5.00$ & $0.39042$ & $-26.89$ & $-25.45$ & $21.096$ & $+22.82$ & $+23.06$ \\ 
   & ToSDT & $0.082517$ & $-6.27$ & $-6.41$ & $3.0293$ & $+1.05$ & $+0.96$ & $0.49783$ & $-6.77$ & $-4.95$ & $17.716$ & $+3.14$ & $+3.35$ \\ 
   & \fozz  & $0.087730$ & $-0.34$ & $-0.50$ & $3.0029$ & $+0.17$ & $+0.08$ & $0.52287$ & $-2.08$ & $-0.16$ & $17.185$ & $+0.05$ & $+0.25$ \\ 
   & \tozz & $0.087908$ & $-0.14$ & $-0.30$ & $3.0001$ & $+0.08$ & $-0.01$ & $0.52265$ & $-2.12$ & $-0.20$ & $17.168$ & $-0.05$ & $+0.15$ \\ 
   & \tdwf  & $0.088171$ & $+0.16$ & $-0.00$ & $3.0005$ & $+0.09$ & $-0.00$ & $0.52371$ & $-1.93$ & $-0.00$ & $17.143$ & $-0.20$ & $+0.00$ \\ 
   & \itwf & $0.088172$ & $+0.16$ & $-0.00$ & $3.0005$ & $+0.09$ & $+0.00$ & $0.52373$ & $-1.92$ & $-0.00$ & $17.143$ & $-0.20$ & $+0.00$ \\ 
   & Exa$^2$ & $0.088172$ &  & \emph{ref.} & $3.0005$ &  & \emph{ref.} & $0.52373$ &  & \emph{ref.} & $17.143$ &  & \emph{ref.} \\ 
   & Exa & $0.088033$ & \emph{ref.} &  & $2.9978$ & \emph{ref.} &  & $0.53399$ & \emph{ref.} &  & $17.177$ & \emph{ref.} &  \\ 

    \hline
    100 & FoSDT  & $0.035362$ & $-0.93$ & $-0.94$ & $3.2418$ & $+0.06$ & $+0.06$ & $0.29572$ & $-0.60$ & $-0.56$ & $27.275$ & $+0.47$ & $+0.47$ \\ 
    & ToSDT & $0.035631$ & $-0.18$ & $-0.18$ & $3.2404$ & $+0.01$ & $+0.01$ & $0.29705$ & $-0.15$ & $-0.12$ & $27.172$ & $+0.09$ & $+0.09$ \\ 
    & \fozz  & $0.035691$ & $-0.01$ & $-0.01$ & $3.2400$ & $+0.00$ & $+0.00$ & $0.29738$ & $-0.05$ & $-0.01$ & $27.149$ & $+0.00$ & $+0.01$ \\ 
    & \tozz & $0.035694$ & $-0.00$ & $-0.01$ & $3.2400$ & $+0.00$ & $+0.00$ & $0.29738$ & $-0.04$ & $-0.00$ & $27.148$ & $+0.00$ & $+0.00$ \\ 
    & \tdwf  & $0.035697$ & $+0.00$ & $-0.00$ & $3.2400$ & $+0.00$ & $+0.00$ & $0.29740$ & $-0.04$ & $-0.00$ & $27.147$ & $-0.00$ & $-0.00$ \\ 
    & \itwf & $0.035697$ & $+0.00$ & $-0.00$ & $3.2400$ & $+0.00$ & $+0.00$ & $0.29740$ & $-0.04$ & $-0.00$ & $27.147$ & $-0.00$ & $-0.00$ \\ 
    & Exa$^2$ & $0.035697$ &  & \emph{ref.} & $3.2400$ &  & \emph{ref.} & $0.29740$ &  & \emph{ref.} & $27.147$ &  & \emph{ref.} \\ 
    & Exa & $0.035695$ & \emph{ref.} &  & $3.2399$ & \emph{ref.} &  & $0.29751$ & \emph{ref.} &  & $27.148$ & \emph{ref.} &  \\ 

	\end{tabular}
	\normalsize
	\caption{Comparison between the different models for the square $[0/c/0]$ sandwich plate with a varying length-to-thickness ratio.}
	\label{tab:Sandwich}
\end{table*}
\subsection{Square $[-15/15]$ antisymmetric angle-ply composite plate}\label{sec:m15p15}
As the two previous cases enter in the cross-ply family, let us now consider an antisymmetric angle-ply square plate with two layers of equal thickness and a $[-15/15]$ stacking sequence. Transverse stresses $\sigma_{23}(0,b/2,z)$ and $\sigma_{13}(a/2,0,z)$ are no longer null for this laminate but $\sigma_{23}(B)$ and $\sigma_{13}(A)$ are. Hence, these stresses have been computed respectively at points $D(0,b/2,h/4)$ and $C(a/2,0,h/4)$ in order to make comparisons between all models. Results are presented in table~\ref{tab:m15p15}. The tendency of predictions is almost the same than for the $[0/90/0]$ case but it can be noticed that poor values are obtained for the stresses at points D and C. These values are influenced by the way the model is able to couple the $x$ and $y$ direction in the kinematic field, \emph{i.e.} the presence of non null $\varphi_{12}(z)$ and $\varphi_{21}(z)$ functions. This is only the case for the \fozz{}, \tozz{}, \tdwf{} and \itwf{} models. However, Cho's model (\tozz{}) does not give very good values for the lowest length-to-thickness values, but it was also the case for the $[0/90/0]$ laminate. That suggests that the problem may be due to other causes; this point is discussed later. The two proposed models, \tdwf{} and \itwf{}, globally give the best results. However, quite poor estimations of transverse shear stresses are obtained at points C and D in the $a/h=2$ case. Comparison with the Exa$^2$ model is let to the section~\ref{sec:discussion}.
\par
Figure~\ref{fig:m15p15_phi} shows the corresponding WF for $a/h=4$, for all plate models except the FoSDT one. Figure~\ref{fig:m15p15_sigma} presents the variations of the transverse shear stresses obtained by integration of the equilibrium equations for all models, compared to the exact solution, in the $a/h=4$ case. 
\iftoggle{submission}{}{\tikzsetnextfilename{m15p15_phi}}
\begin{SmartFigure}[!htb]%
  \centering
  \iftoggle{submission}{
    \includegraphics{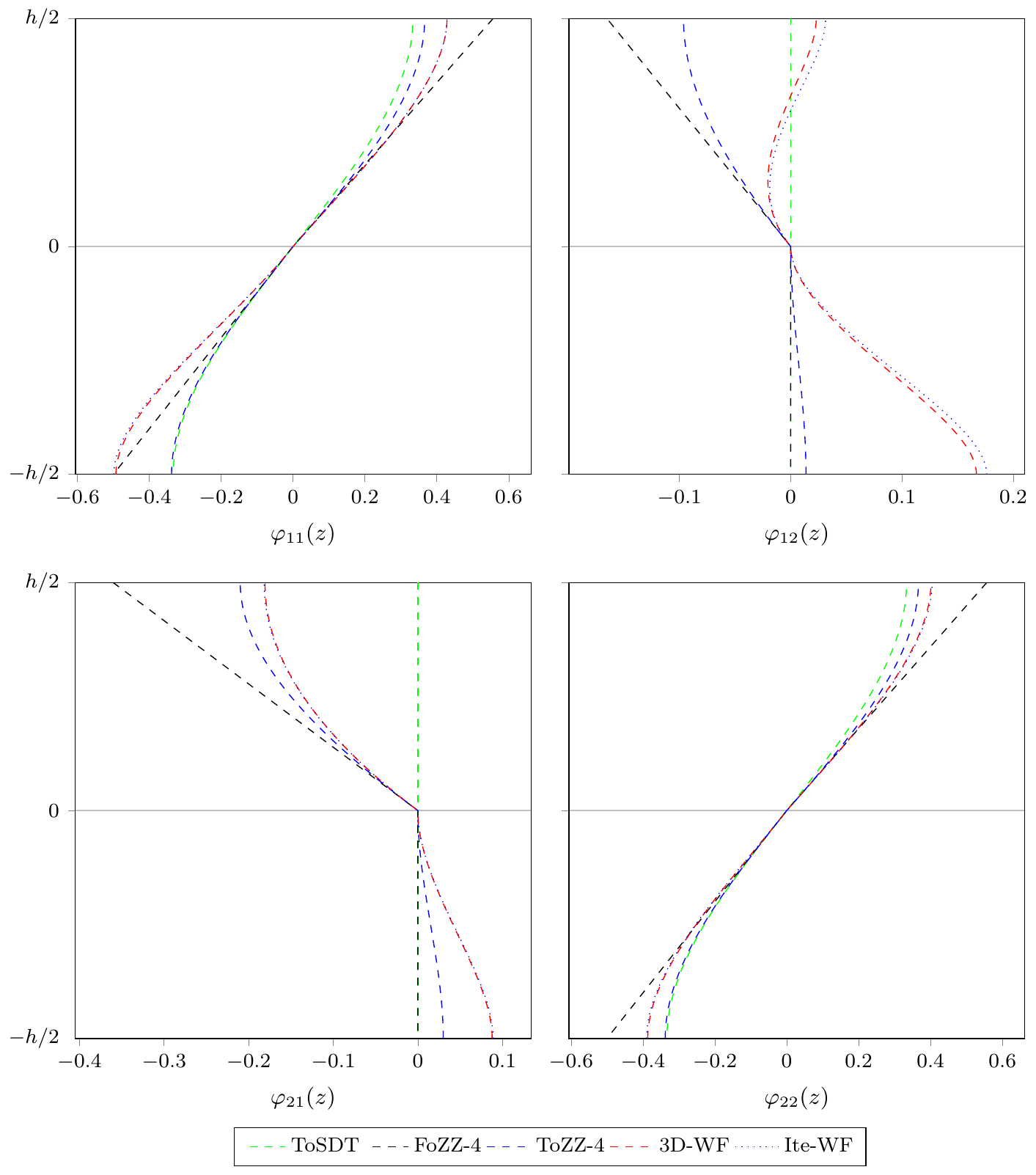}
  }{
  \input{Figures/m15p15_phi.tex}%
  }
  \caption{Transverse shear WF of the $[-15/15]$ antisymmetric angle-ply composite plate with $a/h=4$ for each model.}
  \label{fig:m15p15_phi}
\end{SmartFigure}
\iftoggle{submission}{}{\tikzsetnextfilename{m15p15_sigma}}
\begin{SmartFigure}[!htb]%
  \centering
  \iftoggle{submission}{
    \includegraphics{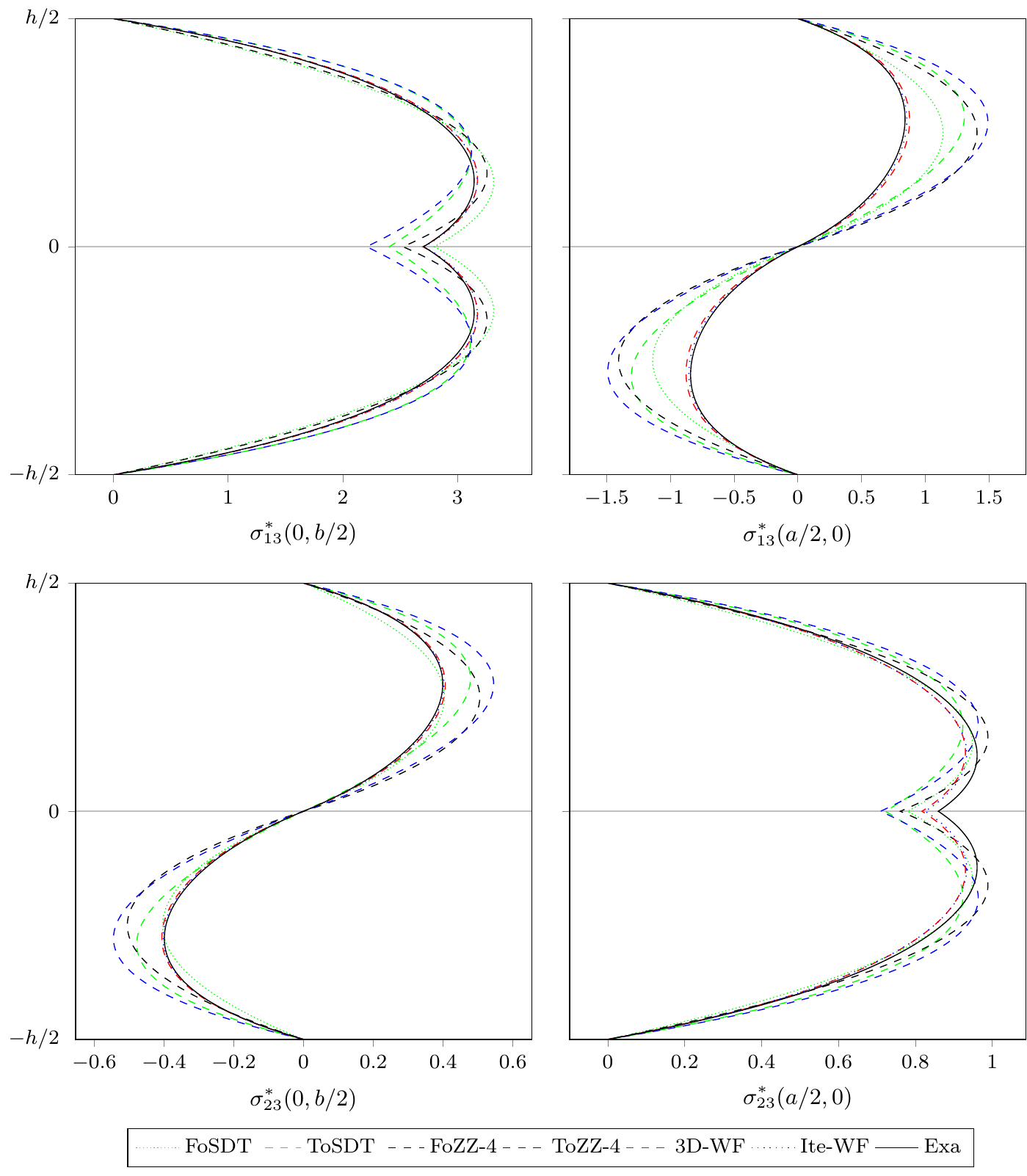}
  }{
  \input{Figures/m15p15_sigma.tex}%
  }
  \caption{Nondimensionalized transverse shear stresses of the $[-15/15]$ antisymmetric angle-ply composite plate with $a/h=4$ for each model.}
  \label{fig:m15p15_sigma}
\end{SmartFigure}
\subsection{Square $[0]$ single ply composite plate}\label{sec:p0}
This square composite plate is made of a single ply oriented at $0^{\circ}$. For this single layer case, the FoSDT model (without shear correction factors) and the \fozz{} exactly coincides. It is also the case for Reddy's and Cho's models (ToSDT and \tozz{}). Note that the $x$ and $y$ directions are not equivalent for the considered laminate, but for Cho's model in this case, $\varphi_{11}(z)=\varphi_{22}(z)$. This is not true for \tdwf{} and \itwf{} models which exhibits different $\varphi_{11}(z)$ and $\varphi_{22}(z)$ functions due to the different longitudinal/shear modulus ratios in each direction. Note also that, even if it has not been presented in this paper, the WF for these two models depend on the length-to-thickness ratio. The proposed models, \tdwf{} and \itwf{}, give poorer values for the deflection than those of the ToSDT model. However, the stresses and the fundamental frequency are best predicted. As we shall see later in section~\ref{sec:discussion}, the plane stress hypothesis is no longer valid for such low length-to-thickness ratios, that is the reason why the Exa$^2$ solution has been introduced.
\par
Figure~\ref{fig:p0_phi} shows the corresponding WF for $a/h=4$, for all plate models except the FoSDT one. Figure~\ref{fig:p0_sigma} presents the variations of the transverse shear stresses obtained by integration of the equilibrium equations for all models, compared to the exact solution, in the $a/h=4$ case. 
\iftoggle{submission}{}{\tikzsetnextfilename{p0_phi}}
\begin{SmartFigure}[!htb]
  \centering
  \iftoggle{submission}{
    \includegraphics{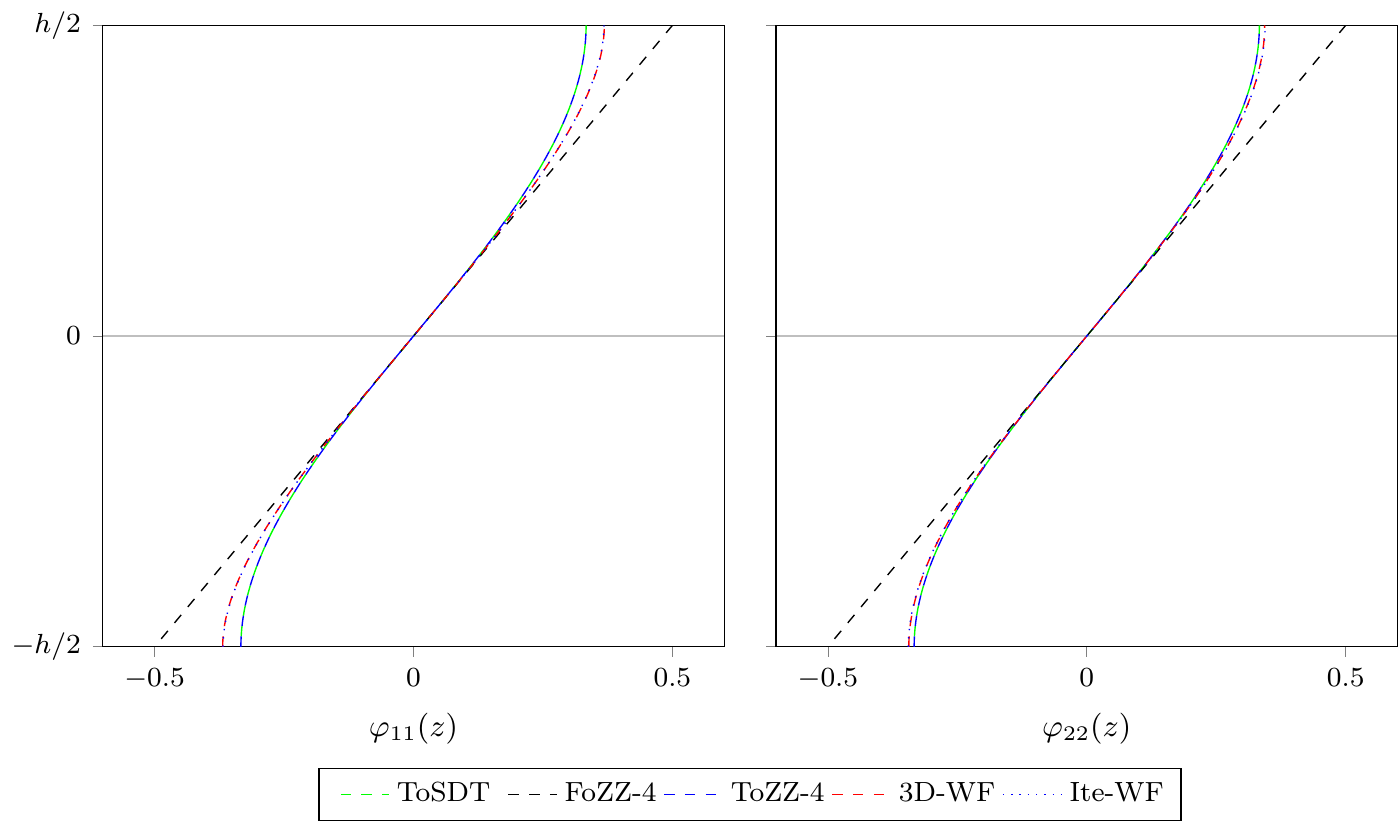}
  }{
    \input{Figures/p0_phi.tex}
  }
  \caption{Transverse shear WF of the $[0]$ single ply composite plate with $a/h=4$ for each model.}%
  \label{fig:p0_phi}
\end{SmartFigure}
\iftoggle{submission}{}{\tikzsetnextfilename{p0_sigma}}
\begin{SmartFigure}[!htb]
  \centering
  \iftoggle{submission}{
    \includegraphics{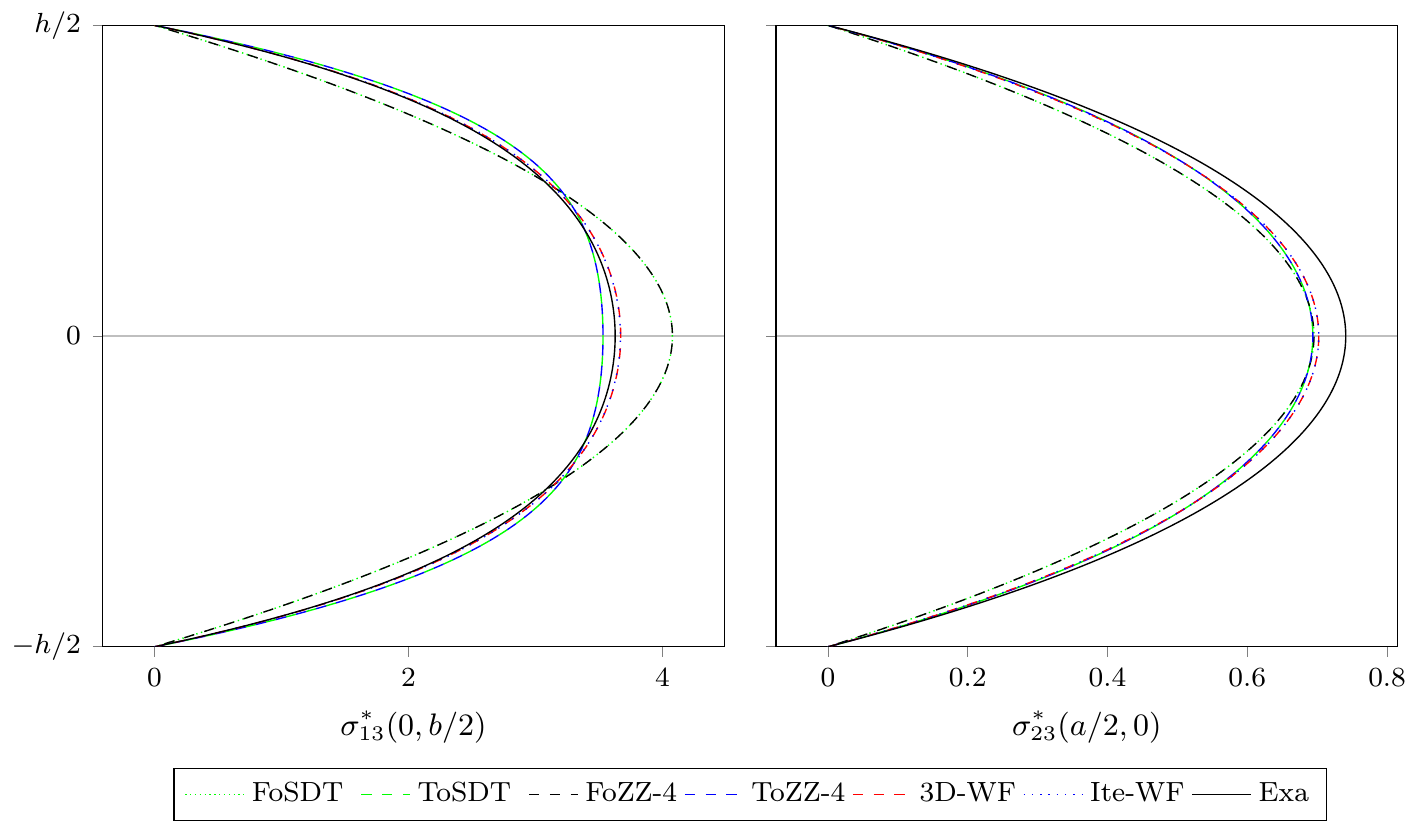}
  }{
    \input{Figures/p0_sigma.tex}
  }
  \caption{Nondimensionalized transverse shear stresses of the $[0]$ single ply composite plate with $a/h=4$ for each model.}%
  \label{fig:p0_sigma}
\end{SmartFigure}
\begin{table*}[!htb]%
	\centering
	\footnotesize
	\begin{tabular}{@{~~}r@{~~}l|l@{~~}r@{~~}r|l@{~~}r@{~~}r|l@{~~}r@{~~}r|l@{~~}r@{~~}r@{~~}}
    a/h & Model & $w^{*}$ & \%~~ & \%~~ & $\sigma_{13}(B)$ & \%~~ & \%~~ & $\sigma_{23}(A)$ & \%~~ & \%~~ & $\omega^{*}$ & \%~~ & \%~~ \\ 

    \hline
    2 & FoSDT  & $4.3108$ & $-3.63$ & $-9.94$ & $3.7502$ & $+31.36$ & $+24.72$ & $1.0244$ & $+2.65$ & $+8.26$ & $4.7281$ & $+4.85$ & $+5.05$ \\ 
  & ToSDT & $4.5262$ & $+1.19$ & $-5.44$ & $2.0291$ & $-28.93$ & $-32.52$ & $0.85739$ & $-14.09$ & $-9.39$ & $4.6223$ & $+2.50$ & $+2.70$ \\ 
  & \fozz  & $4.3108$ & $-3.63$ & $-9.94$ & $3.7502$ & $+31.36$ & $+24.72$ & $1.0244$ & $+2.65$ & $+8.26$ & $4.7281$ & $+4.85$ & $+5.05$ \\ 
  & \tozz & $4.5262$ & $+1.19$ & $-5.44$ & $2.0291$ & $-28.93$ & $-32.52$ & $0.85739$ & $-14.09$ & $-9.39$ & $4.6223$ & $+2.50$ & $+2.70$ \\ 
  & \tdwf  & $4.7804$ & $+6.87$ & $-0.13$ & $3.1758$ & $+11.24$ & $+5.62$ & $0.95157$ & $-4.65$ & $+0.56$ & $4.5040$ & $-0.12$ & $+0.07$ \\ 
  & \itwf & $4.7861$ & $+7.00$ & $-0.01$ & $3.0077$ & $+5.35$ & $+0.03$ & $0.94628$ & $-5.18$ & $+0.00$ & $4.5009$ & $-0.19$ & $+0.00$ \\ 
  & Exa$^2$ & $4.7864$ &  & \emph{ref.} & $3.0069$ &  & \emph{ref.} & $0.94624$ &  & \emph{ref.} & $4.5007$ &  & \emph{ref.} \\ 
  & Exa & $4.4730$ & \emph{ref.} &  & $2.8549$ & \emph{ref.} &  & $0.99796$ & \emph{ref.} &  & $4.5093$ & \emph{ref.} &  \\ 

    \hline
    4 & FoSDT  & $1.4643$ & $-8.42$ & $-10.16$ & $4.0792$ & $+12.44$ & $+11.18$ & $0.69540$ & $-6.14$ & $-0.98$ & $8.1438$ & $+5.17$ & $+5.34$ \\ 
  & ToSDT & $1.6206$ & $+1.35$ & $-0.57$ & $3.5324$ & $-2.63$ & $-3.72$ & $0.69382$ & $-6.35$ & $-1.21$ & $7.7522$ & $+0.11$ & $+0.28$ \\ 
  & \fozz  & $1.4643$ & $-8.42$ & $-10.16$ & $4.0792$ & $+12.44$ & $+11.18$ & $0.69540$ & $-6.14$ & $-0.98$ & $8.1438$ & $+5.17$ & $+5.34$ \\ 
  & \tozz & $1.6206$ & $+1.35$ & $-0.57$ & $3.5324$ & $-2.63$ & $-3.72$ & $0.69382$ & $-6.35$ & $-1.21$ & $7.7522$ & $+0.11$ & $+0.28$ \\ 
  & \tdwf  & $1.6298$ & $+1.93$ & $-0.00$ & $3.6718$ & $+1.21$ & $+0.08$ & $0.70227$ & $-5.21$ & $-0.00$ & $7.7311$ & $-0.16$ & $+0.00$ \\ 
  & \itwf & $1.6298$ & $+1.93$ & $-0.00$ & $3.6693$ & $+1.14$ & $+0.01$ & $0.70231$ & $-5.21$ & $+0.00$ & $7.7311$ & $-0.16$ & $+0.00$ \\ 
  & Exa$^2$ & $1.6299$ &  & \emph{ref.} & $3.6690$ &  & \emph{ref.} & $0.70231$ &  & \emph{ref.} & $7.7310$ &  & \emph{ref.} \\ 
  & Exa & $1.5989$ & \emph{ref.} &  & $3.6280$ & \emph{ref.} &  & $0.74089$ & \emph{ref.} &  & $7.7436$ & \emph{ref.} &  \\ 

    \hline
    10 & FoSDT  & $0.60418$ & $-4.82$ & $-5.17$ & $4.3281$ & $+2.50$ & $+2.32$ & $0.44658$ & $-3.85$ & $-1.85$ & $12.795$ & $+2.59$ & $+2.66$ \\ 
   & ToSDT & $0.63709$ & $+0.37$ & $-0.01$ & $4.2249$ & $+0.06$ & $-0.12$ & $0.45479$ & $-2.08$ & $-0.05$ & $12.464$ & $-0.06$ & $+0.00$ \\ 
   & \fozz  & $0.60418$ & $-4.82$ & $-5.17$ & $4.3281$ & $+2.50$ & $+2.32$ & $0.44658$ & $-3.85$ & $-1.85$ & $12.795$ & $+2.59$ & $+2.66$ \\ 
   & \tozz & $0.63709$ & $+0.37$ & $-0.01$ & $4.2249$ & $+0.06$ & $-0.12$ & $0.45479$ & $-2.08$ & $-0.05$ & $12.464$ & $-0.06$ & $+0.00$ \\ 
   & \tdwf  & $0.63714$ & $+0.37$ & $-0.00$ & $4.2300$ & $+0.18$ & $-0.00$ & $0.45499$ & $-2.03$ & $-0.00$ & $12.463$ & $-0.06$ & $+0.00$ \\ 
   & \itwf & $0.63714$ & $+0.37$ & $-0.00$ & $4.2301$ & $+0.18$ & $+0.00$ & $0.45500$ & $-2.03$ & $-0.00$ & $12.463$ & $-0.06$ & $+0.00$ \\ 
   & Exa$^2$ & $0.63715$ &  & \emph{ref.} & $4.2300$ &  & \emph{ref.} & $0.45500$ &  & \emph{ref.} & $12.463$ &  & \emph{ref.} \\ 
   & Exa & $0.63477$ & \emph{ref.} &  & $4.2223$ & \emph{ref.} &  & $0.46444$ & \emph{ref.} &  & $12.471$ & \emph{ref.} &  \\ 

    \hline
    100 & FoSDT  & $0.43300$ & $-0.08$ & $-0.08$ & $4.3973$ & $+0.03$ & $+0.02$ & $0.37735$ & $-0.06$ & $-0.03$ & $15.196$ & $+0.04$ & $+0.04$ \\ 
    & ToSDT & $0.43335$ & $+0.00$ & $-0.00$ & $4.3962$ & $+0.00$ & $-0.00$ & $0.37746$ & $-0.03$ & $-0.00$ & $15.190$ & $-0.00$ & $-0.00$ \\ 
    & \fozz  & $0.43300$ & $-0.08$ & $-0.08$ & $4.3973$ & $+0.03$ & $+0.02$ & $0.37735$ & $-0.06$ & $-0.03$ & $15.196$ & $+0.04$ & $+0.04$ \\ 
    & \tozz & $0.43335$ & $+0.00$ & $-0.00$ & $4.3962$ & $+0.00$ & $-0.00$ & $0.37746$ & $-0.03$ & $-0.00$ & $15.190$ & $-0.00$ & $-0.00$ \\ 
    & \tdwf  & $0.43335$ & $+0.00$ & $-0.00$ & $4.3962$ & $+0.00$ & $+0.00$ & $0.37746$ & $-0.03$ & $-0.00$ & $15.190$ & $-0.00$ & $+0.00$ \\ 
    & \itwf & $0.43335$ & $+0.00$ & $-0.00$ & $4.3962$ & $+0.00$ & $+0.00$ & $0.37746$ & $-0.03$ & $-0.00$ & $15.190$ & $-0.00$ & $+0.00$ \\ 
    & Exa$^2$ & $0.43335$ &  & \emph{ref.} & $4.3962$ &  & \emph{ref.} & $0.37746$ &  & \emph{ref.} & $15.190$ &  & \emph{ref.} \\ 
    & Exa & $0.43333$ & \emph{ref.} &  & $4.3961$ & \emph{ref.} &  & $0.37756$ & \emph{ref.} &  & $15.190$ & \emph{ref.} &  \\ 

	\end{tabular}
	\normalsize
	\caption{Comparison between the different models for the square $[0]$ single ply composite plate with a varying length-to-thickness ratio.}
	\label{tab:p0}
\end{table*}
\subsection{Square $[0/30/0]$ composite plate}\label{sec:p0p30p0}
Let us now consider a symmetric angle-ply square plate with three layer of equal thickness and a $[0/30/0]$ stacking sequence. This configuration is chosen since it doesn't involve any simplification in the linear system~\eqref{eq:syslin}, i. e. the~$\mymat{K}$ matrix doesn't have any null coefficient. Although this laminate does not represent the more general case (because of its symmetry) it involves an additional $\overline{q}^{mn}\cos(\xi x)\cos(\eta y)$ term in the loading to fulfill the simply supported condition. In other words, it is the only laminate considered in this study which requires a (non-null) Lagrange multiplier in the solving process presented in section~\ref{sec:Navier}. Results presented in table~\ref{tab:p0p30p0} show the same tendency than for the $[0]$ laminate, which in fact is not so different. Present models, \tdwf{} and \itwf{}, give globally better stresses and fundamental frequency, and a poorer deflection, compared to ToSDT or \tozz{} model when the Exa solution is taken as reference.
\par
Figure~\ref{fig:p0p30p0_phi} shows the corresponding WF for $a/h=4$, for all plate models except the FoSDT one. Figure~\ref{fig:p0p30p0_sigma} presents the variations of the transverse shear stresses obtained by integration of the equilibrium equations for all models, compared to the exact solution, in the $a/h=4$ case. 
\iftoggle{submission}{}{\tikzsetnextfilename{p0p30p0_phi}}
\begin{SmartFigure}[!htb]
  \centering
  \iftoggle{submission}{
    \includegraphics{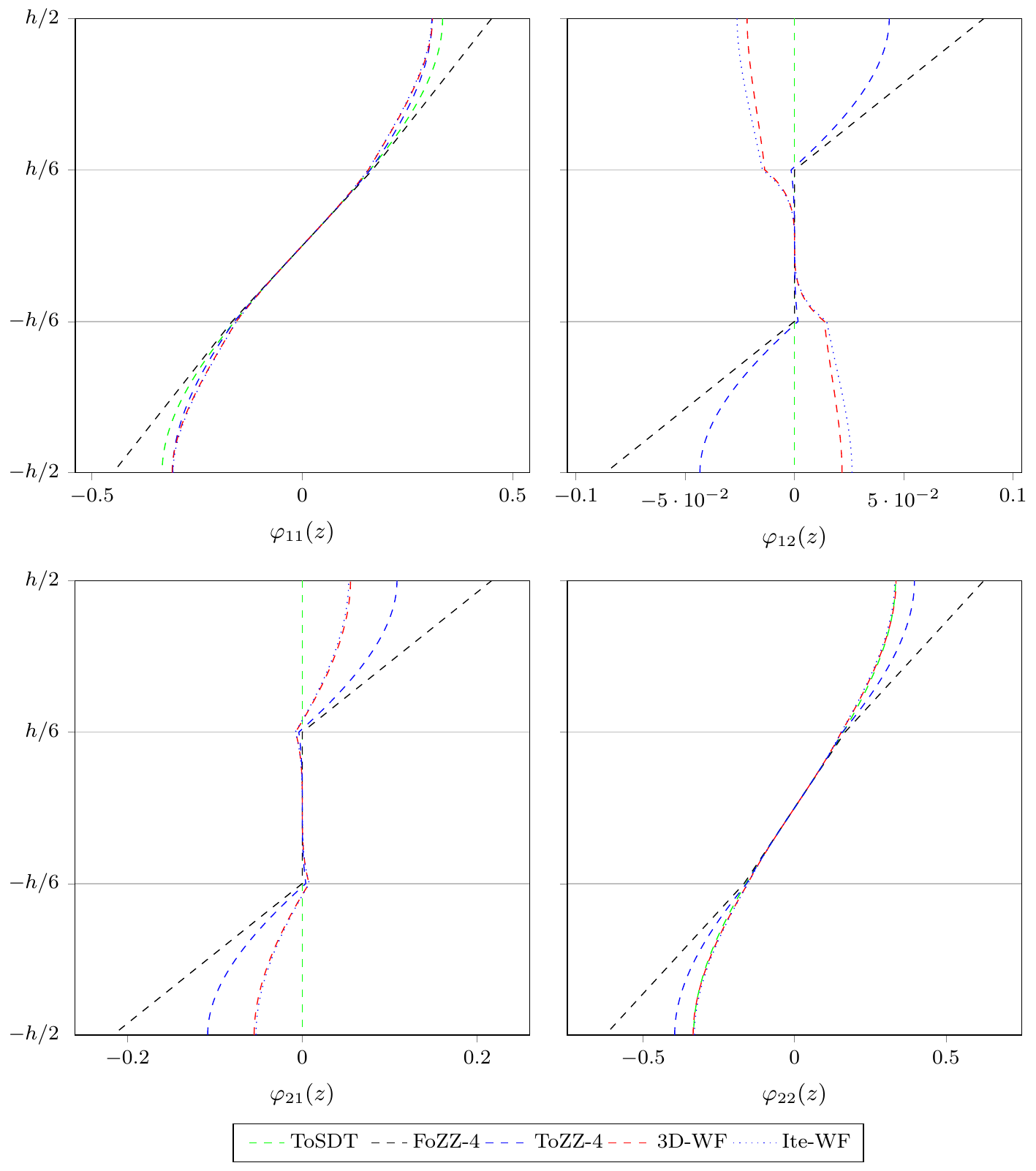}
  }{
    \input{Figures/p0p30p0_phi.tex}
  }  
  \caption{Transverse shear WF of the $[0/30/0]$ composite plate with $a/h=4$ for each model.}%
  \label{fig:p0p30p0_phi}
\end{SmartFigure}
\iftoggle{submission}{}{\tikzsetnextfilename{p0p30p0_sigma}}
\begin{SmartFigure}[!htb]
  \centering
  \iftoggle{submission}{
    \includegraphics{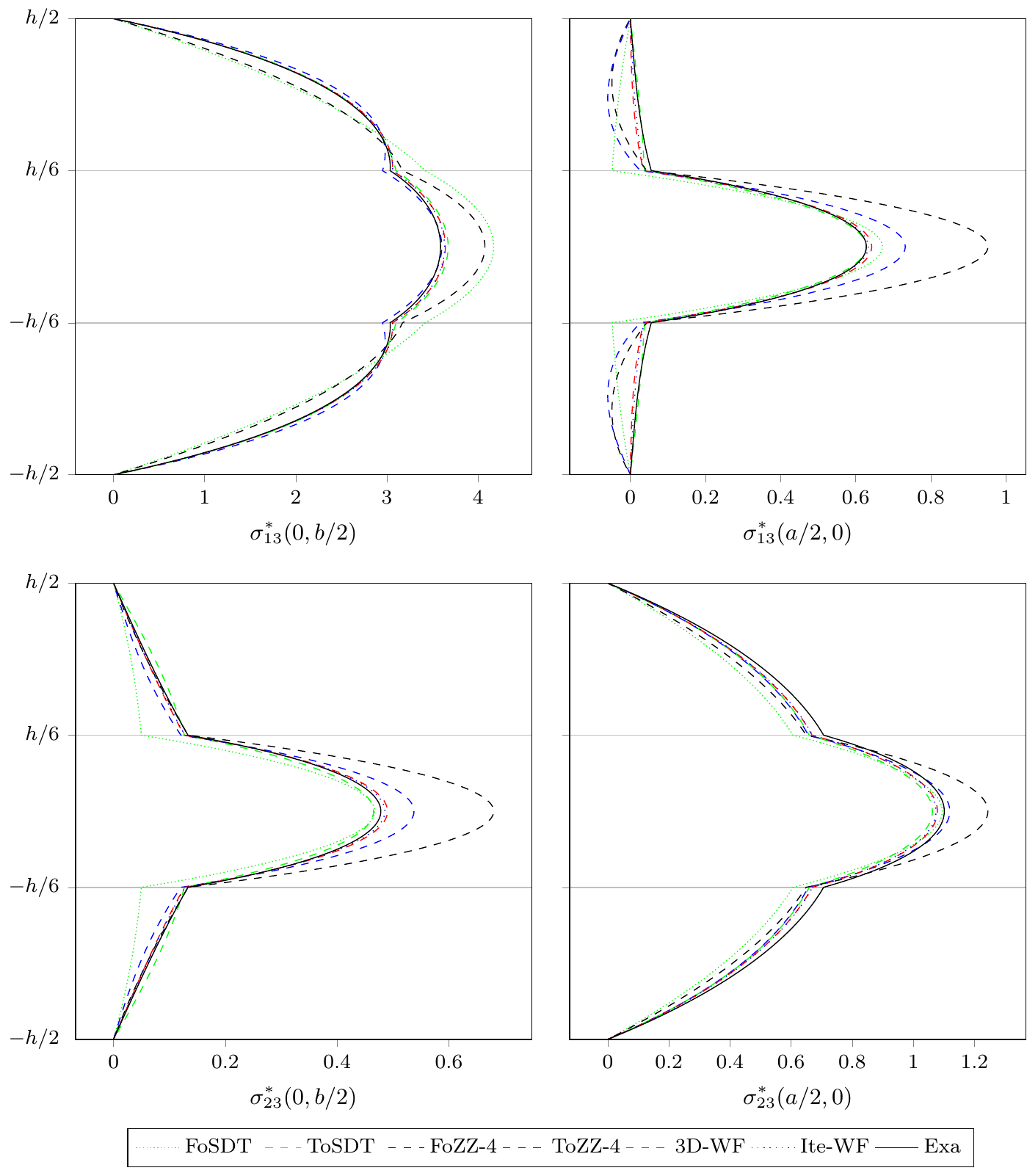}
  }{
    \input{Figures/p0p30p0_sigma.tex}
  }  
  \caption{Nondimensionalized transverse shear stresses of the $[0/30/0]$ composite plate with $a/h=4$ for each model.}%
  \label{fig:p0p30p0_sigma}
\end{SmartFigure}
\subsection{Discussion}\label{sec:discussion}
The five studied cases have been chosen in order to represent a wide range of the laminates diversity. In addition to particular comments done for each case, general results may be issued from these studies. We first compare all models with the Exa analytical solution. The Mindlin model (FoSDT), without the help of appropriate shear correction factors, gives relatively poor values, except for thin plates. For moderately thick plates, Reddy's model (ToSDT) is very accurate, except for the sandwich case because the high modulus ratio between layers is not properly managed by the model. On the contrary, the Sun~\&~Whitney zig-zag linear model (\fozz{}) can manage it but fails for classical laminates, especially the single ply. Cho's model (\tozz{}) which \emph{a priori} takes both advantage of a cubic variation coupled to a zig-zag variation of displacements sometimes gives poor values, especially for thick or very-thick plates. These poor estimations are probably due to the fact that the WF of the \tozz{} model are too constrained by their formulation: as the cubic term is unique for all the layers, the WF sometimes adopt shapes that are not pertinent. In addition, its WF do not depend on the length-to-thickness ratio. Both proposed models, \tdwf{} and \itwf{}, give globally better results in all studied cases than other models. However, some values of the deflection are no so good compared to simpler models. These unexpected results have led to an examination of what was going wrong, and finally to the proposal of the Exa$^2$ solution, presented in section~\ref{sec:Exa2}. The nature and purpose of this Exa$^2$ solution, and the way to obtain it, are there explained. 
\par
The comparison for all models can now be done taking the Exa$^2$ exact solution as reference. It clearly shows that both proposed \tdwf{} and \itwf{} models surpass all other models, giving values for deflection and transverse stresses very close to those of the exact solution, for all studied cases, and for all length to-thickness ratios. The \itwf{} model gives the best results. Of course these spectacular results have to be replaced in the special context of this study. Plate problems solved in this study are simply supported problems with double-sine loading. The WF used for the \tdwf{} and \itwf{} model are obtained considering this special bending problem. Results show that the generic model is able to manage all lamination schemes if appropriate WF are provided, but there is no guarantee that the WF obtained for this special bending configuration will work as well as for other configurations. The model with these WF may be a good candidate for other configurations, compared to simpler models, but other WF could perhaps be considered. Further studies need to be done in order to validate --or adapt-- the WF to other configurations.  
\par
For fundamental frequencies, for both \tdwf{} and \itwf{} models, the WF of the static case have been used. This means that the analytical solution used to obtain the WF for the \tdwf{} model is obtained with $\omega=0$. This $\omega=0$ condition is also used when the successive equilibrium equations are integrated for the \itwf{} model. Although it is possible, without any additional effort, to set in those processes $\omega$ to the appropriate value (given for example by the Exa solution), it has not be done. This seems to have little effect on the WF for frequency values near the fundamental frequency. This can be seen on their shapes (not presented here) and is corroborated by the satisfactory results for fundamental frequencies given in tables. However, one may think that for higher frequencies the WF need to be modified. Hence further studies have to be done with emphasis for the dynamic behavior.
\section{Conclusion}
In this paper, a multilayered equivalent-single-layer plate theory based on warping functions (WF) has been presented. Two ways to generate the WF from three dimensional elasticity equations are described. The first one, leading to the \tdwf{} model, derives directly the WF from a 3D exact solution, called Exa. The second one, leading to the \itwf{} model, is an iterative process involving successive integrations of the equilibrium equations. Both processes use the same core procedure which issues the WF from the variations of transverse shear stresses. Needed variations of the shear stresses are computed at the middle of the edges of the plate, for a simply supported bending problem with a double-sine loading.
\par
These two models are compared to other models and to the exact solution for five lamination sequences. Considered laminated plates are: a $[0/90/0]$ cross-ply rectangular plate, a square $[0/c/0]$ sandwich plate, a square $[0]$ single layer plate, a square $[-15/15]$ antisymmetric angle-ply plate, and a square plate with a more general $[0/30/0]$ lamination sequence. Models for comparison include Mindlin (FoSDT), Reddy (ToSDT), Sun~\&~Whitney--Woodcock (\fozz{}), and Kim-Cho (\tozz{}) models. They have been formulated in terms of WF, so the solution procedure is unique. The problem which is solved for comparison is the simply supported plate with a double-sine loading, for which exact solutions are known. The comparisons are made on deflections, stresses, and fundamental frequencies for length-to-thickness ratios varying from $2$ to $100$. The Mindlin (FoSDT) and Reddy (ToSDT) models are not material-dependent, in other words, their WF are unique. In addition, $\varphi_{11}(z)=\varphi_{22}(z)$ and they never have non-null $\varphi_{12}(z)$ or $\varphi_{21}(z)$ WF. All other considered models have WF which depend on material properties and on the lamination sequence, hence have $4$ different WF for general lamination sequences. In addition, the WF of the \tdwf{} and \itwf{} models depend on the length-to-thickness ratio.
\par
Results show that for classical laminates -- with moderate modulus ratio between layers -- Reddy's model (ToSDT), which has cubic WF, is a good choice for thin and moderately thick plates, but is not so accurate for sandwich structures. On the contrary, the Sun~\&~Whitney--Woodcock model (\fozz{}), which has 4 linear zig-zag WF, is a good choice for the sandwich structure but is not accurate enough for classical laminates. Kim-Cho's model (\tozz{}), which combines advantages from the two previous models, has 4 cubic zig-zag WF. Its results are then better than those of the previous models when we look at all the classes of laminates. However, Kim-Cho's model sometimes gives poor results for low length-to-thickness values, probably because the WF are too constrained by their formulation. Comparisons also show that the two presented models, \tdwf{} and \itwf{}, give globally better results than other tested models for all considered lamination sequences.
\par
For the thick plates ($a/h\le 4$), none of the models can be considered as accurate, even if the \tdwf{} and \itwf{} models give better values on stresses and fundamental frequencies than other models. As all presented models consider the generalized plane stress hypothesis, an exact plane-stress solution called Exa$^2$ has been computed for all considered laminates, in order to make further comparisons. This has been done using the same 3D solution procedure after replacing stiffnesses of all materials with corresponding plane stress stiffnesses. Comparison with the Exa$^2$ reference clearly shows that the \tdwf{} and \itwf{} models give better results than other models. It can be seen that the \itwf{} model gives nearly exact values of deflection, stresses and fundamental frequencies, for all laminates and for all length-to-thickness ratio values.
\par
These last results show clearly that the comparison of considered models with the Exa exact solution for low length-to-thickness values have no sense because these models do not take into account the transverse deformation and use the generalized plane stress hypothesis. This fact has to be related to the notion of consistency as mentioned in the introduction. From another point of view, all these models try --with more or less success-- to simulate another solution which is here denoted Exa$^2$. Hence, comparison of considered models with the Exa solution will led to erroneous conclusions on the ability of these models to integrate the transverse shear behavior. As a result, the transverse deformation must be considered if thick plates have to be studied with equivalent-single-layer theories. It might also be the case for laminates which have high stiffness ratios between layers, and for dynamic studies when the wavelength shortens.  
\par
Results also show that the generation of WF from an analytical solution or from the integration of equilibrium equations is a relatively easy way to have a very accurate universal model, at least for reasonable length-to-thickness ratios. Further studies must try to establish if the WF obtained by these methods are also pertinent for problems involving other boundary conditions, loading, or frequencies. If it is not the case, other ways to compute pertinent WF have to be investigated. 
\begin{landscape}
\begin{table}[p]%
	\centering
	\footnotesize
	\begin{tabular}{@{~}r@{~}l|l@{~}r@{~}r|l@{~}r@{~}r|l@{~}r@{~}r|l@{~}r@{~}r|l@{~}r@{~}r|l@{~}r@{~}r@{~}}
    a/h & Model & $w^{*}$ & \%~~ & \%~~ & $\sigma_{13}(B)$ & \%~~ & \%~~ & $\sigma_{23}(A)$ & \%~~ & \%~~ & $\sigma_{23}(D)$ & \%~~ & \%~~ & $\sigma_{13}(C)$ & \%~~ & \%~~ & $\omega^{*}$ & \%~~ & \%~~ \\ 

    \hline
    2 & FoSDT  & $4.3448$ &  $-4.61$ & $-10.46$ & $2.5852$  & $+10.45$ &  $+9.02$ & $0.96150$ &  $-8.99$ &  $-4.02$ & $0.39006$ &   $+6.66$ &   $+2.00$ & $1.2538$ & $+132.50$ & $+103.37$ & $4.7206$ & $+5.62$ & $+5.70$ \\ 
  & ToSDT & $4.3111$ &  $-5.35$ & $-11.15$ & $1.2945$  & $-44.70$ & $-45.41$ & $0.66314$ & $-37.23$ & $-33.80$ & $0.61086$ &  $+67.03$ &  $+59.73$ & $1.7362$ & $+221.94$ & $+181.61$ & $4.7611$ & $+6.52$ & $+6.61$ \\ 
  & \fozz  & $3.8609$ & $-15.24$ & $-20.43$ & $1.5621$  & $-33.26$ & $-34.13$ & $0.82808$ & $-21.62$ & $-17.34$ & $0.77361$ & $+111.53$ & $+102.29$ & $2.3751$ & $+340.42$ & $+285.24$ & $4.9993$ & $+11.85$ & $+11.94$ \\ 
  & \tozz & $3.9838$ & $-12.54$ & $-17.90$ & $0.68465$ & $-70.75$ & $-71.13$ & $0.58472$ & $-44.65$ & $-41.63$ & $0.85309$ & $+133.26$ & $+123.07$ & $2.4671$ & $+357.48$ & $+300.16$ & $4.9478$ & $+10.70$ & $+10.79$ \\ 
  & \tdwf  & $4.8406$ &  $+6.28$ &  $-0.24$ & $2.3183$   & $-0.96$ &  $-2.24$ & $0.93896$ & $-11.12$ &  $-6.27$ & $0.41381$ &  $+13.15$ &   $+8.21$ & $0.85886$ & $+59.26$ &  $+39.31$ & $4.4826$ & $+0.29$ & $+0.37$ \\ 
  & \itwf & $4.8523$ &  $+6.53$ &  $-0.00$ & $2.3720$   & $+1.34$ &  $+0.03$ & $1.0020$  &  $-5.15$ &  $+0.03$ & $0.38223$ &   $+4.51$ &   $-0.05$ & $0.61566$ & $+14.16$ &   $-0.14$ & $4.4706$ & $+0.03$ & $+0.10$ \\ 
  & Exa$^2$ & $4.8524$ &  & \emph{ref.}  & $2.3713$ &  & \emph{ref.} &       $1.0018$ &  & \emph{ref.}       & $0.38243$ &  & \emph{ref.}        & $0.61652$ &  & \emph{ref.}       & $4.4660$ &  & \emph{ref.} \\ 
  & Exa & $4.5548$ & \emph{ref.} &  &      $2.3407$ & \emph{ref.} &  &       $1.0565$ & \emph{ref.} &        & $0.36572$ & \emph{ref.} &         & $0.53928$ & \emph{ref.} &        & $4.4695$ & \emph{ref.} &  \\ 

    \hline
    4 & FoSDT  & $1.5762$ & $-7.60$ & $-9.49$ & $2.7968$ & $+3.47$ & $+3.29$ & $0.77953$ & $-9.30$ & $-5.53$ & $0.40374$ & $+2.37$ & $+1.35$ & $1.1382$ & $+36.90$ & $+34.50$ & $7.7953$ & $+4.57$ & $+4.88$ \\ 
  & ToSDT & $1.6594$ & $-2.73$ & $-4.72$ & $2.4047$ & $-11.04$ & $-11.19$ & $0.72164$ & $-16.03$ & $-12.55$ & $0.46519$ & $+17.96$ & $+16.78$ & $1.2815$ & $+54.14$ & $+51.43$ & $7.6323$ & $+2.38$ & $+2.69$ \\ 
  & \fozz  & $1.5082$ & $-11.59$ & $-13.40$ & $2.5237$ & $-6.63$ & $-6.79$ & $0.76061$ & $-11.50$ & $-7.83$ & $0.50488$ & $+28.02$ & $+26.74$ & $1.4053$ & $+69.03$ & $+66.06$ & $7.9690$ & $+6.90$ & $+7.22$ \\ 
  & \tozz & $1.6197$ & $-5.05$ & $-7.00$ & $2.2063$ & $-18.38$ & $-18.52$ & $0.70919$ & $-17.48$ & $-14.06$ & $0.53761$ & $+36.32$ & $+34.96$ & $1.4786$ & $+77.85$ & $+74.72$ & $7.7273$ & $+3.66$ & $+3.97$ \\ 
  & \tdwf  & $1.7414$ & $+2.08$ & $-0.01$ & $2.6980$ & $-0.19$ & $-0.36$ & $0.81709$ & $-4.93$ & $-0.98$ & $0.40205$ & $+1.95$ & $+0.93$ & $0.86867$ & $+4.48$ & $+2.65$ & $7.4376$ & $-0.23$ & $+0.07$ \\ 
  & \itwf & $1.7416$ & $+2.09$ & $-0.00$ & $2.7078$ & $+0.17$ & $+0.00$ & $0.82521$ & $-3.98$ & $+0.00$ & $0.39835$ & $+1.01$ & $-0.00$ & $0.84627$ & $+1.79$ & $-0.00$ & $7.4354$ & $-0.26$ & $+0.04$ \\ 
  & Exa$^2$ & $1.7416$ &  & \emph{ref.} & $2.7077$ &  & \emph{ref.} & $0.82520$ &  & \emph{ref.} & $0.39836$ &  & \emph{ref.} & $0.84627$ &  & \emph{ref.} & $7.4324$ &  & \emph{ref.} \\ 
  & Exa & $1.7059$ & \emph{ref.} &  & $2.7031$ & \emph{ref.} &  & $0.85945$ & \emph{ref.} &  & $0.39437$ & \emph{ref.} &  & $0.83139$ & \emph{ref.} &  & $7.4548$ & \emph{ref.} &  \\ 

    \hline
    10 & FoSDT  & $0.77629$ & $-3.29$ & $-3.77$ & $2.9520$ & $+0.67$ & $+0.62$ & $0.64607$ & $-3.11$ & $-1.97$ & $0.41376$ & $+0.54$ & $+0.38$ & $1.0534$ & $+6.32$ & $+6.18$ & $11.218$ & $+1.74$ & $+1.87$ \\ 
   & ToSDT & $0.79522$ & $-0.93$ & $-1.42$ & $2.8811$ & $-1.74$ & $-1.79$ & $0.64072$ & $-3.92$ & $-2.79$ & $0.42393$ & $+3.02$ & $+2.85$ & $1.0791$ & $+8.92$ & $+8.77$ & $11.093$ & $+0.60$ & $+0.73$ \\ 
   & \fozz  & $0.76841$ & $-4.27$ & $-4.75$ & $2.9075$ & $-0.84$ & $-0.89$ & $0.64468$ & $-3.32$ & $-2.18$ & $0.43019$ & $+4.54$ & $+4.36$ & $1.0933$ & $+10.34$ & $+10.19$ & $11.275$ & $+2.26$ & $+2.39$ \\ 
   & \tozz & $0.79198$ & $-1.34$ & $-1.82$ & $2.8463$ & $-2.93$ & $-2.98$ & $0.64029$ & $-3.98$ & $-2.85$ & $0.43619$ & $+6.00$ & $+5.82$ & $1.1098$ & $+12.01$ & $+11.86$ & $11.116$ & $+0.81$ & $+0.94$ \\ 
   & \tdwf  & $0.80669$ & $+0.49$ & $-0.00$ & $2.9334$ & $+0.04$ & $-0.01$ & $0.65877$ & $-1.21$ & $-0.05$ & $0.41233$ & $+0.20$ & $+0.03$ & $0.99287$ & $+0.21$ & $+0.07$ & $11.012$ & $-0.13$ & $+0.00$ \\ 
   & \itwf & $0.80669$ & $+0.49$ & $-0.00$ & $2.9337$ & $+0.05$ & $+0.00$ & $0.65908$ & $-1.16$ & $-0.00$ & $0.41220$ & $+0.17$ & $+0.00$ & $0.99215$ & $+0.13$ & $+0.00$ & $11.012$ & $-0.13$ & $+0.00$ \\ 
   & Exa$^2$ & $0.80669$ &  & \emph{ref.} & $2.9337$ &  & \emph{ref.} & $0.65908$ &  & \emph{ref.} & $0.41220$ &  & \emph{ref.} & $0.99214$ &  & \emph{ref.} & $11.012$ &  & \emph{ref.} \\ 
   & Exa & $0.80272$ & \emph{ref.} &  & $2.9323$ & \emph{ref.} &  & $0.66683$ & \emph{ref.} &  & $0.41152$ & \emph{ref.} &  & $0.99081$ & \emph{ref.} &  & $11.026$ & \emph{ref.} &  \\ 

    \hline
    100 & FoSDT  & $0.62204$ & $-0.04$ & $-0.05$ & $2.9945$ & $+0.01$ & $+0.01$ & $0.60954$ & $-0.04$ & $-0.02$ & $0.41651$ & $+0.01$ & $+0.00$ & $1.0302$ & $+0.06$ & $+0.06$ & $12.677$ & $+0.02$ & $+0.03$ \\ 
    & ToSDT & $0.62224$ & $-0.01$ & $-0.02$ & $2.9937$ & $-0.02$ & $-0.02$ & $0.60950$ & $-0.04$ & $-0.03$ & $0.41661$ & $+0.03$ & $+0.03$ & $1.0305$ & $+0.09$ & $+0.09$ & $12.675$ & $+0.01$ & $+0.01$ \\ 
    & \fozz  & $0.62197$ & $-0.06$ & $-0.06$ & $2.9940$ & $-0.01$ & $-0.01$ & $0.60953$ & $-0.04$ & $-0.03$ & $0.41667$ & $+0.05$ & $+0.04$ & $1.0306$ & $+0.10$ & $+0.10$ & $12.678$ & $+0.03$ & $+0.03$ \\ 
    & \tozz & $0.62222$ & $-0.02$ & $-0.02$ & $2.9934$ & $-0.03$ & $-0.03$ & $0.60950$ & $-0.04$ & $-0.03$ & $0.41673$ & $+0.06$ & $+0.06$ & $1.0308$ & $+0.12$ & $+0.12$ & $12.675$ & $+0.01$ & $+0.01$ \\ 
    & \tdwf  & $0.62235$ & $+0.01$ & $-0.00$ & $2.9943$ & $+0.00$ & $+0.00$ & $0.60969$ & $-0.01$ & $-0.00$ & $0.41649$ & $+0.00$ & $+0.00$ & $1.0296$ & $+0.00$ & $+0.00$ & $12.674$ & $-0.00$ & $+0.00$ \\ 
    & \itwf & $0.62235$ & $+0.01$ & $-0.00$ & $2.9943$ & $+0.00$ & $+0.00$ & $0.60969$ & $-0.01$ & $-0.00$ & $0.41649$ & $+0.00$ & $+0.00$ & $1.0296$ & $+0.00$ & $+0.00$ & $12.674$ & $-0.00$ & $+0.00$ \\ 
    & Exa$^2$ & $0.62235$ &  & \emph{ref.} & $2.9943$ &  & \emph{ref.} & $0.60969$ &  & \emph{ref.} & $0.41649$ &  & \emph{ref.} & $1.0296$ &  & \emph{ref.} & $12.674$ &  & \emph{ref.} \\ 
    & Exa & $0.62232$ & \emph{ref.} &  & $2.9942$ & \emph{ref.} &  & $0.60978$ & \emph{ref.} &  & $0.41648$ & \emph{ref.} &  & $1.0296$ & \emph{ref.} &  & $12.674$ & \emph{ref.} &  \\ 

	\end{tabular}
	\normalsize
	\caption{Comparison between the different models for the square $[-15/15]$ composite plate with a varying length-to-thickness ratio.}
	\label{tab:m15p15}
\end{table}
\end{landscape}
%
%
%
%
%
%
%
%
\begin{landscape}
\begin{table}[p]%
	\centering
	\footnotesize
	\begin{tabular}{@{~}r@{~}l|l@{~}r@{~}r|l@{~}r@{~}r|l@{~}r@{~}r|l@{~}r@{~}r|l@{~}r@{~}r|l@{~}r@{~}r@{~}}
    a/h & Model & $w^{*}$ & \%~~ & \%~~ & $\sigma_{13}(B)$ & \%~~ & \%~~ & $\sigma_{23}(A)$ & \%~~ & \%~~ & $\sigma_{23}(B)$ & \%~~ & \%~~ & $\sigma_{13}(A)$ & \%~~ & \%~~ & $\omega^{*}$ & \%~~ & \%~~ \\ 

    \hline
    2 & FoSDT  & $4.3307$ & $-7.30$ & $-13.36$ & $3.9829$ & $+46.15$ & $+38.02$ & $1.5415$ & $+15.29$ & $+15.38$ & $0.61354$ & $+16.99$ & $+8.40$ & $0.87873$ & $+32.75$ & $+24.61$ & $4.7242$ & $+7.21$ & $+7.47$ \\ 
  & ToSDT & $4.7450$ & $+1.57$ & $-5.07$ & $2.3244$ & $-14.71$ & $-19.45$ & $1.1334$ & $-15.23$ & $-15.16$ & $0.39166$ & $-25.32$ & $-30.80$ & $0.48255$ & $-27.10$ & $-31.57$ & $4.4994$ & $+2.11$ & $+2.35$ \\ 
  & \fozz  & $4.4049$ & $-5.71$ & $-11.87$ & $3.8213$ & $+40.22$ & $+32.42$ & $1.9621$ & $+46.75$ & $+46.87$ & $1.1894$ & $+126.78$ & $+110.13$ & $1.7529$ & $+164.81$ & $+148.57$ & $4.6683$ & $+5.94$ & $+6.19$ \\ 
  & \tozz & $4.5986$ & $-1.56$ & $-8.00$ & $2.3230$ & $-14.76$ & $-19.50$ & $1.3101$ & $-2.02$ & $-1.94$ & $0.57580$ & $+9.79$ & $+1.73$ & $0.90351$ & $+36.49$ & $+28.12$ & $4.5753$ & $+3.83$ & $+4.08$ \\ 
  & \tdwf  & $4.9868$ & $+6.74$ & $-0.23$ & $3.1129$ & $+14.22$ & $+7.87$ & $1.4246$ & $+6.55$ & $+6.64$ & $0.64957$ & $+23.86$ & $+14.76$ & $0.86657$ & $+30.91$ & $+22.89$ & $4.4067$ & $+0.01$ & $+0.24$ \\ 
  & \itwf & $4.9982$ & $+6.99$ & $-0.00$ & $2.8858$ & $+5.89$ & $+0.00$ & $1.3360$ & $-0.08$ & $+0.00$ & $0.56602$ & $+7.93$ & $+0.00$ & $0.70518$ & $+6.53$ & $-0.00$ & $4.3977$ & $-0.20$ & $+0.04$ \\ 
  & Exa$^2$ & $4.9983$ &  & \emph{ref.} & $2.8857$ &  & \emph{ref.} & $1.3360$ &  & \emph{ref.} & $0.56601$ &  & \emph{ref.} & $0.70519$ &  & \emph{ref.} & $4.3960$ &  & \emph{ref.} \\ 
  & Exa & $4.6717$ & \emph{ref.} &  & $2.7253$ & \emph{ref.} &  & $1.3370$ & \emph{ref.} &  & $0.52445$ & \emph{ref.} &  & $0.66195$ & \emph{ref.} &  & $4.4064$ & \emph{ref.} &  \\ 

    \hline
    4 & FoSDT  & $1.4728$ & $-13.63$ & $-15.24$ & $4.1710$ & $+16.24$ & $+14.82$ & $1.0945$ & $-0.61$ & $+1.89$ & $0.46704$ & $-2.34$ & $-3.79$ & $0.67093$ & $+6.79$ & $+5.95$ & $8.1258$ & $+8.40$ & $+8.58$ \\ 
  & ToSDT & $1.7025$ & $-0.15$ & $-2.02$ & $3.6717$ & $+2.32$ & $+1.07$ & $1.0624$ & $-3.53$ & $-1.11$ & $0.46523$ & $-2.72$ & $-4.16$ & $0.62730$ & $-0.16$ & $-0.94$ & $7.5603$ & $+0.86$ & $+1.03$ \\ 
  & \fozz  & $1.5616$ & $-8.42$ & $-10.13$ & $4.0752$ & $+13.57$ & $+12.18$ & $1.2440$ & $+12.96$ & $+15.80$ & $0.68005$ & $+42.20$ & $+40.09$ & $0.95180$ & $+51.49$ & $+50.31$ & $7.8819$ & $+5.15$ & $+5.32$ \\ 
  & \tozz & $1.7138$ & $+0.51$ & $-1.37$ & $3.6017$ & $+0.37$ & $-0.85$ & $1.1193$ & $+1.63$ & $+4.19$ & $0.53754$ & $+12.40$ & $+10.74$ & $0.73172$ & $+16.46$ & $+15.55$ & $7.5351$ & $+0.52$ & $+0.69$ \\ 
  & \tdwf  & $1.7376$ & $+1.90$ & $-0.00$ & $3.6411$ & $+1.47$ & $+0.23$ & $1.0788$ & $-2.05$ & $+0.42$ & $0.49005$ & $+2.47$ & $+0.95$ & $0.64235$ & $+2.24$ & $+1.44$ & $7.4846$ & $-0.15$ & $+0.01$ \\ 
  & \itwf & $1.7376$ & $+1.91$ & $-0.00$ & $3.6327$ & $+1.24$ & $+0.00$ & $1.0743$ & $-2.45$ & $+0.00$ & $0.48544$ & $+1.51$ & $+0.00$ & $0.63325$ & $+0.79$ & $+0.00$ & $7.4841$ & $-0.16$ & $+0.01$ \\ 
  & Exa$^2$ & $1.7376$ &  & \emph{ref.} & $3.6327$ &  & \emph{ref.} & $1.0743$ &  & \emph{ref.} & $0.48543$ &  & \emph{ref.} & $0.63324$ &  & \emph{ref.} & $7.4836$ &  & \emph{ref.} \\ 
  & Exa & $1.7051$ & \emph{ref.} &  & $3.5883$ & \emph{ref.} &  & $1.1013$ & \emph{ref.} &  & $0.47822$ & \emph{ref.} &  & $0.62828$ & \emph{ref.} &  & $7.4961$ & \emph{ref.} &  \\ 

    \hline
    10 & FoSDT  & $0.59918$ & $-8.29$ & $-8.63$ & $4.3112$ & $+3.04$ & $+2.85$ & $0.74022$ & $-2.55$ & $-1.48$ & $0.32843$ & $-4.09$ & $-4.20$ & $0.51166$ & $-0.21$ & $-0.30$ & $12.850$ & $+4.51$ & $+4.58$ \\ 
   & ToSDT & $0.64701$ & $-0.97$ & $-1.34$ & $4.2213$ & $+0.89$ & $+0.70$ & $0.75108$ & $-1.12$ & $-0.03$ & $0.34034$ & $-0.62$ & $-0.73$ & $0.51673$ & $+0.78$ & $+0.69$ & $12.370$ & $+0.61$ & $+0.67$ \\ 
   & \fozz  & $0.62084$ & $-4.98$ & $-5.33$ & $4.2884$ & $+2.49$ & $+2.30$ & $0.77032$ & $+1.41$ & $+2.53$ & $0.37135$ & $+8.44$ & $+8.32$ & $0.57195$ & $+11.55$ & $+11.45$ & $12.625$ & $+2.68$ & $+2.75$ \\ 
   & \tozz & $0.65454$ & $+0.18$ & $-0.19$ & $4.1991$ & $+0.36$ & $+0.17$ & $0.76235$ & $+0.36$ & $+1.47$ & $0.35476$ & $+3.59$ & $+3.47$ & $0.53317$ & $+3.98$ & $+3.89$ & $12.299$ & $+0.03$ & $+0.10$ \\ 
   & \tdwf  & $0.65579$ & $+0.37$ & $-0.00$ & $4.1920$ & $+0.19$ & $+0.00$ & $0.75140$ & $-1.08$ & $+0.01$ & $0.34297$ & $+0.15$ & $+0.04$ & $0.51344$ & $+0.14$ & $+0.05$ & $12.287$ & $-0.06$ & $+0.00$ \\ 
   & \itwf & $0.65579$ & $+0.37$ & $-0.00$ & $4.1919$ & $+0.19$ & $+0.00$ & $0.75131$ & $-1.09$ & $+0.00$ & $0.34285$ & $+0.12$ & $+0.00$ & $0.51321$ & $+0.09$ & $+0.00$ & $12.287$ & $-0.06$ & $+0.00$ \\ 
   & Exa$^2$ & $0.65579$ &  & \emph{ref.} & $4.1919$ &  & \emph{ref.} & $0.75131$ &  & \emph{ref.} & $0.34285$ &  & \emph{ref.} & $0.51320$ &  & \emph{ref.} & $12.287$ &  & \emph{ref.} \\ 
   & Exa & $0.65337$ & \emph{ref.} &  & $4.1841$ & \emph{ref.} &  & $0.75958$ & \emph{ref.} &  & $0.34245$ & \emph{ref.} &  & $0.51275$ & \emph{ref.} &  & $12.295$ & \emph{ref.} &  \\ 

    \hline
    100 & FoSDT  & $0.42339$ & $-0.14$ & $-0.14$ & $4.3503$ & $+0.03$ & $+0.03$ & $0.63890$ & $-0.04$ & $-0.03$ & $0.28614$ & $-0.07$ & $-0.07$ & $0.47133$ & $-0.01$ & $-0.01$ & $15.367$ & $+0.07$ & $+0.07$ \\ 
    & ToSDT & $0.42390$ & $-0.02$ & $-0.02$ & $4.3494$ & $+0.01$ & $+0.01$ & $0.63907$ & $-0.02$ & $-0.00$ & $0.28630$ & $-0.01$ & $-0.01$ & $0.47142$ & $+0.01$ & $+0.01$ & $15.358$ & $+0.01$ & $+0.01$ \\ 
    & \fozz  & $0.42363$ & $-0.08$ & $-0.09$ & $4.3501$ & $+0.03$ & $+0.02$ & $0.63922$ & $+0.01$ & $+0.02$ & $0.28660$ & $+0.09$ & $+0.09$ & $0.47203$ & $+0.14$ & $+0.14$ & $15.363$ & $+0.04$ & $+0.04$ \\ 
    & \tozz & $0.42399$ & $+0.00$ & $-0.00$ & $4.3491$ & $+0.00$ & $+0.00$ & $0.63919$ & $+0.00$ & $+0.02$ & $0.28646$ & $+0.04$ & $+0.04$ & $0.47159$ & $+0.05$ & $+0.05$ & $15.356$ & $+0.00$ & $+0.00$ \\ 
    & \tdwf  & $0.42400$ & $+0.00$ & $-0.00$ & $4.3490$ & $+0.00$ & $+0.00$ & $0.63907$ & $-0.01$ & $+0.00$ & $0.28633$ & $+0.00$ & $+0.00$ & $0.47137$ & $+0.00$ & $+0.00$ & $15.356$ & $-0.00$ & $+0.00$ \\ 
    & \itwf & $0.42400$ & $+0.00$ & $-0.00$ & $4.3490$ & $+0.00$ & $+0.00$ & $0.63907$ & $-0.01$ & $+0.00$ & $0.28633$ & $+0.00$ & $+0.00$ & $0.47137$ & $+0.00$ & $+0.00$ & $15.356$ & $-0.00$ & $+0.00$ \\ 
    & Exa$^2$ & $0.42400$ &  & \emph{ref.} & $4.3490$ &  & \emph{ref.} & $0.63907$ &  & \emph{ref.} & $0.28633$ &  & \emph{ref.} & $0.47137$ &  & \emph{ref.} & $15.356$ &  & \emph{ref.} \\ 
    & Exa & $0.42398$ & \emph{ref.} &  & $4.3490$ & \emph{ref.} &  & $0.63917$ & \emph{ref.} &  & $0.28633$ & \emph{ref.} &  & $0.47136$ & \emph{ref.} &  & $15.356$ & \emph{ref.} &  \\ 

	\end{tabular}
	\normalsize
	\caption{Comparison between the different models for the square $[0/30/0]$ composite plate with a varying length-to-thickness ratio.}
	\label{tab:p0p30p0}
\end{table}
\end{landscape}

\bibliographystyle{model3-num-names}
\bibliography{TwoMultPlateModWithTSWF3D}
\appendix
\section{Appendices}
\subsection{Detailed formulation of the stiffness and mass matrices}
\label{sec:detailedmatrices}
\scriptsize
\begin{landscape}
\begin{align*} 
  &
    K=\left[\begin{array}{cc}
      K_1 & K_2 \\
      K_3 & K_4
    \end{array}\right]
    \qquad
    \text{and}
    \qquad
      M=\left[\begin{array}{cc}
        M_1 & M_2 \\
        M_3 & M_4
      \end{array}\right]
    \qquad
    \text{with}
  \\
  &
K_1=
\left[
\begin{array}{ccccc}
-A_{1212}\eta^2-A_{1111}\xi^2 & -A_{1122}\xi\eta-A_{1212}\xi\eta & B_{1122}\xi\eta^2+2B_{1212}\xi\eta^2+B_{1111}{\xi}^{3} & -E_{1111}\xi^2-E_{1221}\eta^2 & -E_{1112}\xi\eta-E_{1222}\xi\eta\\
-A_{1122}\xi\eta-A_{1212}\xi\eta & -A_{2222}\eta^2-A_{1212}\xi^2 & B_{2222}{\eta}^{3}+B_{1122}\xi^2\eta+2B_{1212}\xi^2\eta & -E_{2211}\xi\eta-E_{1221}\xi\eta & -E_{1222}\xi^2-E_{2212}\eta^2\\
B_{1122}\xi\eta^2+2B_{1212}\xi\eta^2+B_{1111}{\xi}^{3} & B_{2222}{\eta}^{3}+B_{1122}\xi^2\eta+2B_{1212}\xi^2\eta & -D_{1111}{\xi}^{4}-4D_{1212}\xi^2\eta^2-D_{2222}{\eta}^{4}-2D_{1122}\xi^2\eta^2 & 2F_{1221}\xi\eta^2+F_{1111}{\xi}^{3}+F_{2211}\xi\eta^2 & F_{1112}\xi^2\eta+2F_{1222}\xi^2\eta+F_{2212}{\eta}^{3}\\
-E_{1111}\xi^2-E_{1221}\eta^2 & -E_{2211}\xi\eta-E_{1221}\xi\eta & 2F_{1221}\xi\eta^2+F_{1111}{\xi}^{3}+F_{2211}\xi\eta^2 & -G_{1111}\xi^2-G_{2121}\eta^2-H_{1212} & -G_{1112}\xi\eta-G_{2122}\xi\eta\\
-E_{1112}\xi\eta-E_{1222}\xi\eta & -E_{1222}\xi^2-E_{2212}\eta^2 & F_{1112}\xi^2\eta+2F_{1222}\xi^2\eta+F_{2212}{\eta}^{3} & -G_{1112}\xi\eta-G_{2122}\xi\eta & -G_{2222}\xi^2-G_{1212}\eta^2-H_{2323}
\end{array}
\right]
\\
  &
K_2=
\left[
\begin{array}{ccccc}
-2A_{1112}\xi\eta & -A_{2212}\eta^2-A_{1112}\xi^2 & -B_{2212}{\eta}^{3}-3B_{1112}\xi^2\eta & -E_{1211}\xi\eta-E_{1121}\xi\eta & -E_{1122}\xi^2-E_{1212}\eta^2\\
-A_{2212}\eta^2-A_{1112}\xi^2 & -2A_{2212}\xi\eta & -B_{1112}{\xi}^{3}-3B_{2212}\xi\eta^2 & -E_{2221}\eta^2-E_{1211}\xi^2 & -E_{1212}\xi\eta-E_{2222}\xi\eta\\
3B_{1112}\xi^2\eta+B_{2212}{\eta}^{3} & 3B_{2212}\xi\eta^2+B_{1112}{\xi}^{3} & 4D_{1112}{\xi}^{3}\eta+4D_{2212}\xi{\eta}^{3} & F_{2221}{\eta}^{3}+2F_{1211}\xi^2\eta+F_{1121}\xi^2\eta & F_{1122}{\xi}^{3}+F_{2222}\xi\eta^2+2F_{1212}\xi\eta^2\\
-E_{1211}\xi\eta-E_{1121}\xi\eta & -E_{2221}\eta^2-E_{1211}\xi^2 & -2F_{1211}\xi^2\eta-F_{2221}{\eta}^{3}-F_{1121}\xi^2\eta & -2G_{1121}\xi\eta & -G_{1221}\eta^2-H_{1213}-G_{1122}\xi^2\\
-E_{1122}\xi^2-E_{1212}\eta^2 & -E_{1212}\xi\eta-E_{2222}\xi\eta & -2F_{1212}\xi\eta^2-F_{1122}{\xi}^{3}-F_{2222}\xi\eta^2 & -G_{1221}\eta^2-H_{1213}-G_{1122}\xi^2 & -2G_{1222}\xi\eta
\end{array}
\right]
\\
  &
K_3=
\left[
\begin{array}{ccccc}
-2A_{1112}\xi\eta & -A_{2212}\eta^2-A_{1112}\xi^2 & 3B_{1112}\xi^2\eta+B_{2212}{\eta}^{3} & -E_{1211}\xi\eta-E_{1121}\xi\eta & -E_{1122}\xi^2-E_{1212}\eta^2\\
-A_{2212}\eta^2-A_{1112}\xi^2 & -2A_{2212}\xi\eta & 3B_{2212}\xi\eta^2+B_{1112}{\xi}^{3} & -E_{2221}\eta^2-E_{1211}\xi^2 & -E_{1212}\xi\eta-E_{2222}\xi\eta\\
-B_{2212}{\eta}^{3}-3B_{1112}\xi^2\eta & -B_{1112}{\xi}^{3}-3B_{2212}\xi\eta^2 & 4D_{1112}{\xi}^{3}\eta+4D_{2212}\xi{\eta}^{3} & -2F_{1211}\xi^2\eta-F_{2221}{\eta}^{3}-F_{1121}\xi^2\eta & -2F_{1212}\xi\eta^2-F_{1122}{\xi}^{3}-F_{2222}\xi\eta^2\\
-E_{1211}\xi\eta-E_{1121}\xi\eta & -E_{2221}\eta^2-E_{1211}\xi^2 & F_{2221}{\eta}^{3}+2F_{1211}\xi^2\eta+F_{1121}\xi^2\eta & -2G_{1121}\xi\eta & -G_{1221}\eta^2-H_{1213}-G_{1122}\xi^2\\
-E_{1122}\xi^2-E_{1212}\eta^2 & -E_{1212}\xi\eta-E_{2222}\xi\eta & F_{1122}{\xi}^{3}+F_{2222}\xi\eta^2+2F_{1212}\xi\eta^2 & -G_{1221}\eta^2-H_{1213}-G_{1122}\xi^2 & -2G_{1222}\xi\eta
\end{array}
\right]
\\
  &
K_4=
\left[
\begin{array}{ccccc}
-A_{1212}\eta^2-A_{1111}\xi^2 & -A_{1122}\xi\eta-A_{1212}\xi\eta & -2B_{1212}\xi\eta^2-B_{1111}{\xi}^{3}-B_{1122}\xi\eta^2 & -E_{1111}\xi^2-E_{1221}\eta^2 & -E_{1112}\xi\eta-E_{1222}\xi\eta\\
-A_{1122}\xi\eta-A_{1212}\xi\eta & -A_{2222}\eta^2-A_{1212}\xi^2 & -B_{1122}\xi^2\eta-B_{2222}{\eta}^{3}-2B_{1212}\xi^2\eta & -E_{2211}\xi\eta-E_{1221}\xi\eta & -E_{1222}\xi^2-E_{2212}\eta^2\\
-2B_{1212}\xi\eta^2-B_{1111}{\xi}^{3}-B_{1122}\xi\eta^2 & -B_{1122}\xi^2\eta-B_{2222}{\eta}^{3}-2B_{1212}\xi^2\eta & -D_{1111}{\xi}^{4}-4D_{1212}\xi^2\eta^2-D_{2222}{\eta}^{4}-2D_{1122}\xi^2\eta^2 & -F_{1111}{\xi}^{3}-F_{2211}\xi\eta^2-2F_{1221}\xi\eta^2 & -F_{1112}\xi^2\eta-F_{2212}{\eta}^{3}-2F_{1222}\xi^2\eta\\
-E_{1111}\xi^2-E_{1221}\eta^2 & -E_{2211}\xi\eta-E_{1221}\xi\eta & -F_{1111}{\xi}^{3}-F_{2211}\xi\eta^2-2F_{1221}\xi\eta^2 & -G_{1111}\xi^2-G_{2121}\eta^2-H_{1212} & -G_{1112}\xi\eta-G_{2122}\xi\eta\\
-E_{1112}\xi\eta-E_{1222}\xi\eta & -E_{1222}\xi^2-E_{2212}\eta^2 & -F_{1112}\xi^2\eta-F_{2212}{\eta}^{3}-2F_{1222}\xi^2\eta & -G_{1112}\xi\eta-G_{2122}\xi\eta & -G_{2222}\xi^2-G_{1212}\eta^2-H_{2323}
\end{array}
\right]
\\
  &
M_1=
\left[
\begin{array}{ccccc}
R & 0 & -S\xi & U_{11} & 0\\
0 & R & -S\eta & 0 & U_{22}\\
-S\xi & -S\eta & R+T\xi^2+T\eta^2 & -V_{11}\xi & -V_{22}\eta\\
U_{11} & 0 & -V_{11}\xi & W_{11} & 0\\
0 & U_{22} & -V_{22}\eta & 0 & W_{22}
\end{array}
\right]
\qquad
M_2=
\left[
\begin{array}{ccccc}
0 & 0 & 0 & 0 & U_{12}\\
0 & 0 & 0 & U_{21} & 0\\
0 & 0 & 0 & -V_{21}\eta & -V_{12}\xi\\
0 & U_{21} & V_{21}\eta & 0 & W_{12}\\
U_{12} & 0 & V_{12}\xi & W_{21} & 0
\end{array}
\right]
\\
  &
M_3=
\left[
\begin{array}{ccccc}
0 & 0 & 0 & 0 & U_{12}\\
0 & 0 & 0 & U_{21} & 0\\
0 & 0 & 0 & V_{21}\eta & V_{12}\xi\\
0 & U_{21} & -V_{21}\eta & 0 & W_{12}\\
U_{12} & 0 & -V_{12}\xi & W_{21} & 0
\end{array}
\right]
\qquad
M_4=
\left[
\begin{array}{ccccc}
R & 0 & S\xi & U_{11} & 0\\
0 & R & S\eta & 0 & U_{22}\\
S\xi & S\eta & R+T(\xi^2+\eta^2) & V_{11}\xi & V_{22}\eta\\
U_{11} & 0 & V_{11}\xi & W_{11} & 0\\
0 & U_{22} & V_{22}\eta & 0 & W_{22}
\end{array}
\right]
\end{align*}
\end{landscape}
\label{lastpage}
\end{document}